\documentclass[%
 reprint,
 superscriptaddress,
 amsmath,amssymb,
 aps,
 longbibliography,
 prb,
floatfix,
]{revtex4-1}

\usepackage{graphicx}% Include figure files
\usepackage{dcolumn}% Align table columns on decimal point
\usepackage{bm}% bold math
\usepackage{natbib,ifthen}
\newcommand{\Kbar}{\bar{\text{K}}}
\newcommand{\Jbar}{\bar{\text{J}}}
\newcommand{\Jpbar}{\bar{\text{J'}}}

\newcommand{\Gbar}{\bar{\Gamma}}

\begin{document}

%\preprint{APS/123-QED}

\title{Electronic properties of Bi-doped GaAs(001) semiconductors}% Force line breaks with \\
\thanks{A footnote to the article title}%

%\author{J. Honolka\textsuperscript{1}, C. D. Hogan\textsuperscript{2}, M. Vondr\'a\v{c}ek\textsuperscript{1}, Y. Polyak\textsuperscript{1}, F. Arciprete\textsuperscript{1} and }
%\address{\textsuperscript{1}Institute of Physics, Academy of Sciences of the Czech Republic, Na Slovance 2, CZ-182 21 Praha 8, Czech Republic}
%\address{Istituto di Struttura della Materia, CNR-ISM, Via del Fosso del Cavaliere 100, 00133 Roma, Italy and Universit\`a di Roma``Tor Vergata'', Via della Ricerca Scientifica 1, 00133 Roma, Italy}

\author{J. Honolka}
\affiliation{Institute of Physics, Academy of Sciences of the
Czech Republic, Na Slovance 2, CZ-182 21 Praha 8, Czech Republic}

\author{C. Hogan}
\affiliation{Istituto di Struttura della Materia-CNR (ISM-CNR), Via del Fosso del Cavaliere 100, 00133 Roma, Italy}
\affiliation{Universit\`{a} di Roma "Tor Vergata", Dipartimento di Fisica, via della Ricerca Scientifica 1, 00133 Roma Italy}

\author{M. Vondr\'a\v{c}ek}  
\affiliation{Institute of Physics, Academy of Sciences of the
Czech Republic, Na Slovance 2, CZ-182 21 Praha 8, Czech Republic}

\author{Y. Polyak}
\affiliation{Institute of Physics, Academy of Sciences of the
Czech Republic, Na Slovance 2, CZ-182 21 Praha 8, Czech Republic}

\author{F. Arciprete}
\affiliation {Universit\`{a} di Roma "Tor Vergata", Dipartimento di Fisica, via della Ricerca Scientifica 1, 00133 Roma Italy} 
\affiliation{Istituto di Struttura della Materia-CNR (ISM-CNR), Via del Fosso del Cavaliere 100, 00133 Roma, Italy}

\author{E. Placidi}
\affiliation{Istituto di Struttura della Materia-CNR (ISM-CNR), Via del Fosso del Cavaliere 100, 00133 Roma, Italy}
\affiliation{Universit\`{a} di Roma "Tor Vergata", Dipartimento di Fisica, via della Ricerca Scientifica 1, 00133 Roma Italy}


%\date{\today \textbf{ Version 1.3}}
\date{\today }

\begin{abstract}
Despite its potential in the fields of optoelectronics and topological insulators, experimental electronic band structure studies of Bi-doped GaAs are scarce. The reason is the complexity of growth which tends to leave bulk and in particular surface properties in an undefined state.
Here we present an in depth investigation of structural and electronic properties of GaAsBi epilayers grown by molecular beam epitaxy with high (001) crystalline order and well-defined surface structures evident from low-energy electron diffraction. X-ray and ultraviolet photoemission spectrocopy as well as angle-resolved photoemission data at variable photon energies allows to disentangle a Bi-rich surface layer with $(1\times3)$ symmetry from the effects of Bi atoms incorporated in the GaAs bulk matrix. The influence of Bi concentrations up to $\approx 1$\% integrated in the GaAs bulk  are visible in angle-resolved photoemission spectra after mild ion bombardment and subsequent annealing steps. Interpretation of our results is obtained via density functional theory simulations of bulk and $\beta 2(2\times 4)$ reconstructed slab geometries with and without Bi. Bi-induced energy shifts in the dispersion of GaAs heavy and light hole bulk bands are evident both in experiment and theory, which are relevant for modulations in the optical band gap and thus optoelectronic applications.
\end{abstract}

\pacs{Valid PACS appear here}% PACS, the Physics and Astronomy
\maketitle

%\tableofcontents
%\section{References}
%Just a placeholder section, add your DOI's here:
%
%Refs.\onlinecite{Olde1990,Bannow2016,Wang2010,Schmidt1996,Gray2012,Abdiche2010,Strasser2001,Bastiman2012,Malone2013,Souma2016,Duzik2014,Achour2008,Gray2011,Larsen1982,Luo2015,Habchi2014,Duzik2012,Chiang1983,Madouri2008,Chuang2014,Cai1992,Pashartis2017,Laukkanen2017,Francoeur2003,marko2016,yamashita2006,richards2015}

%%%%%%%%%%%%%%%%%%%%%%%%%%%%%%%%%%%%%%%%%%%%%%%%%%%%%%%%%%%%%%%%%%%%%%%%%%
\section{Introduction \label{sec:intro}}
%%%%%%%%%%%%%%%%%%%%%%%%%%%%%%%%%%%%%%%%%%%%%%%%%%%%%%%%%%%%%%%%%%%%%%%%%%

GaAs$_{1-x}$Bi$_{x}$ with varying Bi concentrations has enormous potential for a number of advanced applications in the fields of electronics, optoelectronics, nanophotonics, thermoelectricity, photovoltaics \cite{Marko2016,Richards2015,Yamashita2006}. The incorporation of Bi into GaAs induces a large band gap bowing even at Bi concentrations less than few percent~\cite{Francoeur2003}, due to a large upward shift of the valence band edge. This property offers a large freedom to engineer the band structure of the alloy for potential high-speed electronic and infrared optoelectronic applications, even since Bi alloying in GaAs$_{1-x}$Bi$_{x}$ seems to improve the hole mobility, while preserving the high electron mobility of GaAs~\cite{Kini2009}.

Despite the strategic interest of the GaAs$_{1-x}$Bi$_{x}$ alloy, a reliable phase of this material with well controlled properties is still far from having being grown for several reasons. In fact, the existence of a large miscibility gap requires low growth temperatures leading to the formation of defects that degrade optical properties, while the surfactant action of Bi counteracts its incorporation and can leads to phase-separation phenomena (such as surface Ga-Bi droplets). Finally, the large-radius Bi atom leads to a compressive lattice distortion of the GaAs lattice.
To couple the optical/electronic properties of the GaAs$_{1-x}$Bi$_{x}$ alloy and the aggregation properties of Bi in the alloy it is very interesting to study the alloy with traditional and innovative surface techniques. 

To this aim, in this paper we challenge the matter of Bi inclusion in GaAs lattice by means of X-ray Photoemission spectroscopy (XPS) and we study the electronic band structure by means of Ultraviolet Photoemission Spectroscopy (UPS) and $k$-resolved Photoemission Electron Microscopy ($k$-PEEM). By varying the Bi flux during MBE growth we are able to change the Bi concentration in the bulk. Generally, our results show Bi in two different states. One corresponds to a Bi-rich surface layer. The ($1\times 3$) Low Energy Electron Diffraction (LEED) symmetry of the surface suggests the formation of a phase similar to the one found after deposition of Bi onto to $\beta 2(2\times 4)$ GaAs(001) surface, as reported in literature \cite{Bastiman2012}. The second Bi phase corresponds to Bi atoms integrated in the bulk with concentrations up to 1\%, essential for bulk-related applications, e.g. in optics. After removing the Bi-rich surface layer we are able to achieve well-ordered surfaces, which allows to study Bi-concentration dependent band structure properties via UPS and $k$-PEEM. $k$-PEEM and highly-resolved UPS measurements at synchrotron facilities reveal that the integration of Bi into GaAs leads to a broadening of the band structure and a shift in the dispersion of GaAs bulk heavy hole and light hole bands $\Delta_{3,4}$ and $\Delta_1$. Band structure results are interpreted with the help of density functional theory calculations including spin-orbit coupling of the electronic properties of bulk GaAs and $\beta 2(2\times 4)$ reconstructed slab geometries with and without incorporation of Bi. 
%%% here a couple of sentences to introduce results %%%
%%%%%%%

%%%%%%%%%%%%%%%%%%%%%%%%%%%%%%%%%%%%%%%%%%%%%%%%%%%%%%%%%%%%%%%%%%%%%%%%%%
\section{Methodology \label{sec:methodology}}
%%%%%%%%%%%%%%%%%%%%%%%%%%%%%%%%%%%%%%%%%%%%%%%%%%%%%%%%%%%%%%%%%%%%%%%%%%

%=----------------------------------------------------------------------=
\subsection{Sample preparation procedure and experimental methods \label{sec:methods}}
%=----------------------------------------------------------------------=
%%%%%%%%%%%%%%%%%%%%%%%%%%
\begin{figure}[tbh!]
\includegraphics[width=3.0in,height=1.5in]{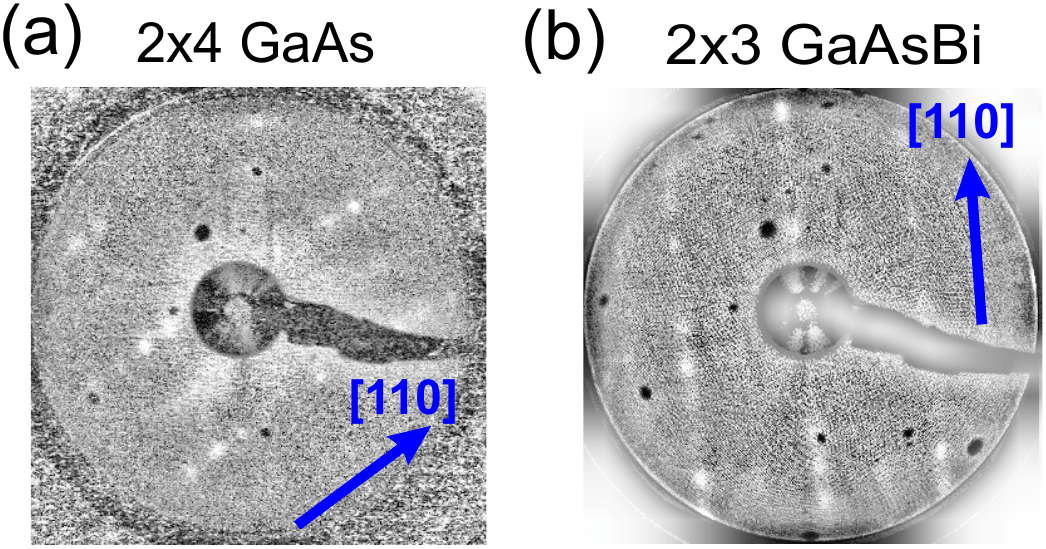}
\caption{ LEED images for (a) GaAs(001)-$\beta_2(2\times4)$ and (b) {\it sputter-annealed} GaAs${_{1-\text{x}}}$Bi$_\text{x}$ (x=2.7\%) with $(2\times 3)$ surface reconstructions. 
\label{fig:manuscript-LEED-image}
}
\end{figure}
%%%%%%%%%%%%%%%%%%%%%%%%%%%

Pure GaAs and GaAsBi samples were grown by MBE on n(Si)-doped (10$^{18}$ cm$^{-3}$) GaAs(001) substrates with a miscut angle of about 0.01$^\circ $. GaAs buffer layers of approximate thickness 500 nm were grown at 590 $^\circ$C. For the growth of GaAsBi layers, the substrate temperature was decreased to 325$^\circ $C. Prior to GaAsBi deposition. A layer of 150 nm of GaAsBi was deposited at a flux rate of 0.5~ML/s.
At the end of both GaAs and GaAsBi epilayer deposition, samples were cooled down to about -10$^\circ$C, and an amorphous As capping layer was deposited to prevent the sample from contamination during air exposure when \emph{ex situ} shipped to XPS and \emph{k}-PEEM facilities.

Prior to surface sensitive photoemission electron microscopy (PEEM), LEED, XPS, and $k$-PEEM measurements, the samples were decapped \emph{in situ} under ultra-high vacuum (UHV) conditions at temperatures of 300$^\circ$C and subsequently annealed at 400-440 $^\circ$C. Temperatures were monitored directly on the sample using an optical pyrometer. Resulting surface order was investigated by LEED where a clear $\beta2(2\times 4)$ pattern is found for pure GaAs(001) as shown in Fig.~\ref{fig:manuscript-LEED-image}(a). On the other hand, GaAsBi samples showed much weaker LEED structures according to a $(1\times 3)$ pattern. We will in the following refer to these samples as {\it decapped}.

As we discuss below, after decapping we observe a Bi-rich phase at the surface of GaAs$_\text{1-x}$Bi$_\text{x}$, with $\text{x}>0$. Since we are particularly interested in the bulk properties we treated GaAsBi samples by a short ion bombardment (up to 12 min) and subsequent annealing to again 400-440 $^\circ$C. In the following we refer to respective samples as {\it sputter-annealed}. Typical LEED images of GaAsBi sputter-annealed samples are shown in Fig.~\ref{fig:manuscript-LEED-image}(b). In the case of GaAsBi sputter-annealed, surfaces exhibit sharper LEED patterns with respect to the respective as-decapped surfaces. A clear $(2\times 3)$ pattern is evident in Fig.~\ref{fig:manuscript-LEED-image}(b).

XPS and \emph{k}-PEEM measurements were carried out by using an \emph{Omicron
NanoESCA} instrument with laboratory light sources. Highly resolved XPS-UPS
was measured independently at the ELETTRA synchrotron radiation facility (MSB beamline). The NanoESCA photoemission spectrometer is based on a PEEM column and an
imaging double hemispherical energy filter \cite{Kromker2008} with an energy resolution of $\Delta E =$ 0.1~eV. A transfer lens in the electron optics switches between the real space and the angle-resolved
\emph{k}-PEEM mode, which allows to detect classical x-ray photoemission
spectra with monochromatized Al $K\alpha$ radiation, as well as an
energy dependent mapping of the Brillouin zone (BZ) using a helium discharge
lamp at $h\nu$ = 21.2~eV. 
%%% is this the resolution of XPS or UPS? %%%
%In the \emph{k}-PEEM mode the Fermi edge $E_{\text F}$ is usually derived from the kinetic energy at which the \emph{k}-PEEM
%intensity is cut off. We define this energy as zero binding energy (BE) and
%expect the error of this Fermi level estimation to be $\pm0.05$~eV.

%=----------------------------------------------------------------------=
\subsection{Theoretical simulations \label{sec:theory}}
%=----------------------------------------------------------------------=
% %%%%%%%%%%%%%%%%%%%%%%%%%%
% \begin{figure}
% %\includegraphics[width=5.0in,height=2.9in]{Images/structure.pdf}
% \caption{(a) Surface reconstruction of GaAs(001)-$\beta 2(2\times4)$. 
% (b) Measured LEED.
% (c) Side view of 2x4 slab, showing surface and bulk regions. Position of As/Bi substitution is indicated. Perhaps also some surface state.
% }
% \label{fig:structure}
% \end{figure}
% %%%%%%%%%%%%%%%%%%%%%%%%%%%

Geometry and electronic structure calculations were carried out using density functional theory (DFT) within the local density approximation (LDA) as implemented in the {\sc quantum ESPRESSO} code\cite{Giannozzi2017}. Spin-orbit coupling was accounted for by means of fully relativistic ultrasoft pseudopotentials,\cite{DalCorso2005} including nonlinear core corrections for all species and semicore d states in the valence for both Ga and Bi.\cite{Rappe1990} 
The kinetic energy cutoff was 30 Ry (300 Ry for the charge), and the theoretical lattice constant of 5.61\AA---slightly smaller than the experimental one of 5.65\AA---was adopted. Structural relaxations used a threshold of 5meV/\AA. 

Two types of system were considered. First, a 128-atom $4\times4\times4$ supercell was used to compute the electronic structure of bulk GaAs and GaAsBi. 
In the latter, Bi was substituted for As at densities of 1.6\% and 3.2\% (1 and 2 Bi atoms per cell, respectively), 
and all atoms were allowed to relax.
$\Gamma$-centred $(2\times2\times2)$ k-point meshes were used. 
Second, the GaAs(001)-$\beta 2(2 \times 4)$ surface reconstruction 
%[see Fig.~\ref{fig:structure}(a)] 
was studied using a supercell geometry. 
Thin GaAs slabs containing ten atomic layers (14\AA\ thick) were separated from their periodic replicas by 20 \AA. 
$\Gamma$-centred $(4\times8\times1)$ k-point meshes were used. The As-terminated back surface was fixed to bulk positions and passivated with pseudohydrogen.
As shown in Section~\ref{sec:XPS}, Bi is mostly present in the bulk layers of the sputter-annealed surface. 
%the  on XPS we wFrom the XPS data in Fig.~\ref{fig:XPS} Guided by the XPS observations of minimal for the sputter-annealed surfaces we study the influence of Bi substitution in the bulk layers only 
%(see Fig.~\ref{fig:structure}(a)). 
%By maintaining the same $2\times4$ $2\times4$ reconstruction, 
%In order to accurately study the influence of %Bi substitution on the bulk and bulk-like %properties of the system, 
We therefore consider the effect of Bi substitution only in the central bulk layers of the $2\times4$ reconstructed slab.
%(see Fig.~\ref{fig:structure}(a)).
Thus we do not consider the measured $1\times3$ periodicity of the GaAsBi surface nor the possible influence of Bi on surface states.  %(Fig.~\ref{fig:structure}(x)), 
Note that the precise $1\times3$ structure is anyway difficult to ascertain, being a mixture of $4\times3$ phases possibly featuring Bi-As or Bi-Bi dimers on the surface\cite{Bastiman2012}.
%DFT calculations have anyway shown that %$4\times3$ reconstructions of pure GaAs are %only $\sim$25 meV higher in energy than the %$\beta_2(2\times4)$ across a wide range of As %chemical potential.\cite{Thomas2010}. 
%Hence it is difficult to say, especially considering the small CL remaining from surface Bi and independence of annealing time, if mixed Bi-As or Bi-Bi dimers are present, and canot be ruled out.
%Hence, we expect our approach to be a reasonable approximation of the true system.
% Figure to show definition of surface.
%Check k-points for relaxation

The computed electronic properties were analysed through different techniques.
First, the band structures of the bulk $4\times4\times4$ and slab $2\times4$ supercells were unfolded along high symmetry directions in the primitive bulk GaAs BZ and $1\times1$ surface BZ (SBZ), respectively, using the BandUP code .\cite{Medeiros2014,Medeiros2015}.
This allow us to make a direct comparison with k-PEEM data and helps decouple surface and Bi-related states from bulk states of GaAs. 
%in order to allow direct comparison with k-PEEM data, using the BandUP code.\cite{Medeiros2014,Medeiros2015} A similar unfolding procedure was performed for bulk GaAs and GaAsBi $4\times4\times4$ supercells along the $\Gamma$--X (100) and $\Gamma$--L (111) lines of the bulk.
Second, the integrated density of states (DOS) and k-resolved DOS of the $2\times4$ slab and of bulk GaAs were computed. In the slab case, the DOS was further projected onto surface and bulk atomic orbitals, respectively, in which the surface is defined as including all dimers and backbonds (layers 1--4), and the bulk makes up the remainder (layers 5--10).
The k-resolved DOS was computed along various high symmetry lines as a function of energy and also as constant energy cuts (2D maps) through the full reduced zone SBZ.
%Fig.~\ref{fig:structure}(x)).
%The integrated density of states (DOS) and k-resolved DOS of the slab were then computed for the whole slab and for projections onto the surface and bulk atomic orbitals, respectively. 
%Second, the band structure was computed along high symmetry lines and unfolded across the $1\times1$ SBZ in order to allow direct comparison with k-PEEM data, using the BandUP code.\cite{Medeiros2014,Medeiros2015} A similar unfolding procedure was performed for bulk GaAs and GaAsBi $4\times4\times4$ supercells along the $\Gamma$--X (100) and $\Gamma$--L (111) lines of the bulk.
%PBBS for 1x1?
%The use of such relatively thin slabs causes bulk states and deep surface resonances to undergo shifts in energy due to quantization of their wavefunctions. In order to simplify comparison with the experimental data and to align properly the projected bulk bandstructures, we perform a simple linear stretching of the slab bands such that the VBM and peak in the DOS at 4eV matched that of the bulk. This may cause a small shift (\textless 0.1 eV) in the position of surface states.
%Size quantization in slab ?
% Name of van hove singularity at 4ev? Looks 2D
 %Later discuss mapping onto surface and bulk directions

%%%%%%%%%%%%%%%%%%%%%%%%%%%%%%%%%%%%%%%%%%%%%%%%%%%%%%%%%%%%%%%%%%%%%%%%%%
\section{X-ray photoemission spectroscopy \label{sec:XPS}}
%%%%%%%%%%%%%%%%%%%%%%%%%%%%%%%%%%%%%%%%%%%%%%%%%%%%%%%%%%%%%%%%%%%%%%%%%%

%%%%%%%%%%%%%%%%%%%%%%%%%%
\begin{figure}[tb!]
\includegraphics[width=0.48\textwidth]{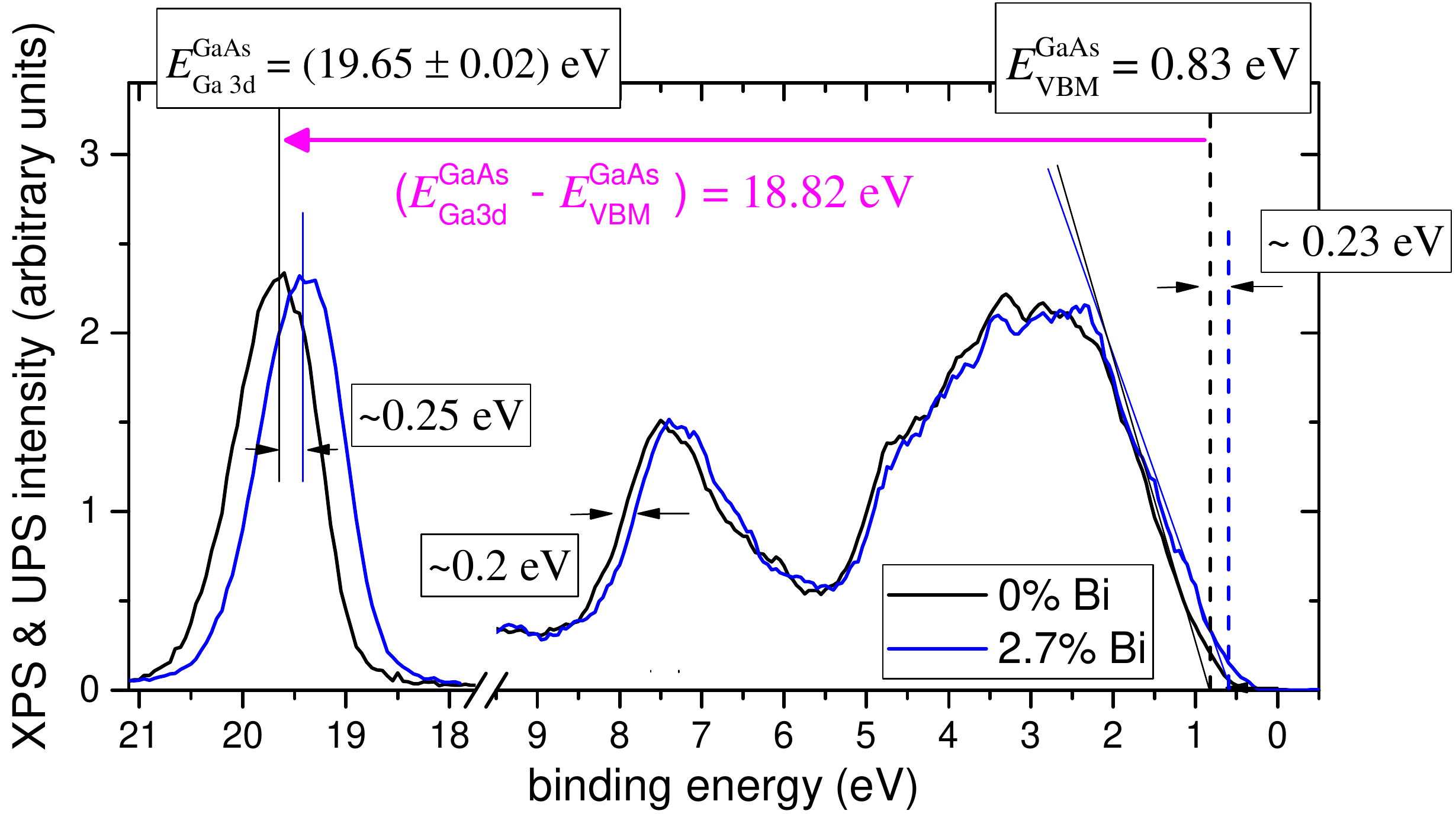}
\caption{Valence band spectra and shallow Ga 3d CLs measured at 1.5~keV. The data of GaAs $\beta 2(2\times 4)$ (black) and GaAs${_{1-\text{x}}}$Bi$_\text{x}$ (x=2.7\%) (blue) after decapping are plotted. Values $E_{\text{VBM}}^{\text{GaAs}}$ and $E_{\text{VBM}}^{\text{GaAsBi}}$ were derived using the leading edge method.
\label{fig:Kraut}
}
\end{figure}
%%%%%%%%%%%%%%%%%%%%%%%%%%%

%%%%%%%%%%%%%%%%%%%%%%%%%%
\begin{figure}[tbp!]
\includegraphics[width=3.3in,height=3.6in]{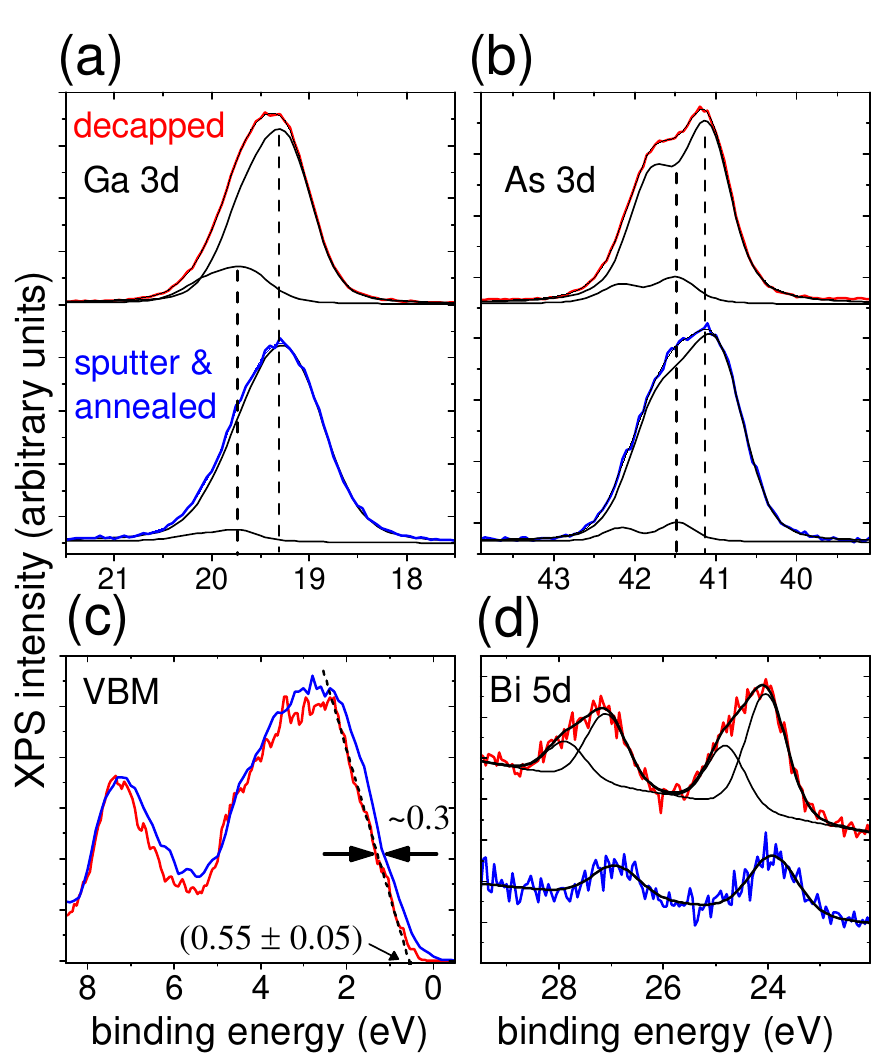}
\caption{Sputtering and Annealing effect on (Ga 3d, Bi 5d, As 3d) CLs (a) and (d), as well as valence band spectra (c) measured at 1.5~keV. GaAs${_{1-\text{x}}}$Bi$_\text{x}$ (x=2.7\%) is shown in its decapped (blue) and sputtered-annealed state (red). The value $E_{\text{VBM}}^{\text{GaAsBi}}=0.55 $~eV in the annealed state is derived using the leading edge method indicated by the tangential dashed line.}
\label{fig:Kraut-sputtering}
\end{figure}
%%%%%%%%%%%%%%%%%%%%%%%%%%%
%%%%%%%%%%%%%%%%%%%%%%%%%%%%%%%%%%
\begin{table}[tb!]
  \begin{tabular}{|l|lll|}
%    \hline
    Bi content & 0 $\%$ & 0.8 $\%$ & 2.7 $\%$\\
    \hline
		XPS [eV] & 0.83 & - & 0.60 \\
%		\hline
        $k$-PEEM [eV] & 0.68 & 0.50 & 0.55\\
%    \hline
    UPS [eV] & 0.75 & 0.35 & 0.34\\
%    \hline		
  \end{tabular}
  \caption{Overview of VBM energy values $E_{\text{VBM}}$ derived from XPS, UPS, $k$-PEEM data shown in Fig.~\ref{fig:Kraut}, Fig.~\ref{fig:VBM-kPEEM}, and Fig.~\ref{fig:UPS-VBM}, respectively. Values were extracted using the Kraut method. We estimate error bars to be 0.05 eV.
  \label{tbl:VBMvalues}
   }
\end{table}
%%%%%%%%%%%%%%%%%%%%%%%%%%%%%%%%%

%=----------------------------------------------------------------------=
%\subsection{Referencing CLs to the valence band maximum \label{sec:VBM}}
%=----------------------------------------------------------------------=

Before we discuss GaAs$_{1-x}$Bi$_{x}$ XPS data we remind the reader that in semiconductors the Fermi level ($E_{\text F}$) with respect to valence band maximum (VBM) and conduction band minimum (CBM) is sensitive to defect states and possible surface band bending effects. We expect Bi dopants in GaAs to cause shifts of $E_{\text F}$, which in turn will affect binding energy (BE) values detected in photoemission experiments. Kraut {\it et al.}\cite{Kraut1980} described procedures to quantify defect induced shifts in $E_{\text F}$ and band bending in GaAs by referencing shallow core level (CL) BEs $E_{\text{Ga 3d} }$ to the VBM value $E_{\text{VBM}}$. Fig.~\ref{fig:Kraut} shows low BE photoemission data of GaAs(001)$(2\times 4)$ in comparison with decapped Bi-doped GaAs measured at $h\nu$ = 1.5 keV. The high photon energies correspond to a probing depth of several nanometers and thus the data reflects mainly bulk properties. We estimate the VBM $E_{\text{VBM}}$ using the leading edge method, which approximates the DOS by a line at the maximum steepness of the VB edge as indicated in Fig.~\ref{fig:Kraut}. 

In the case of GaAs(001)$(2\times 4)$ we derive a value $E_{\text{VBM}}^{\text{GaAs}} = (0.83 \pm 0.02)$~eV. We want to stress at this point that from UPS data discussed below, we have good evidence that the absolute value $E_{\text{VBM}}^{\text{GaAs}}$ does not seem to be affected by band bending effects. The relative BE of the Ga 3d level with respect to VBM then amounts to $(E_{\text{Ga 3d}}^{\text{GaAs}} - E_{\text{VBM}}^{\text{GaAs}}) = (18.82 \pm 0.02)$~eV, which is in good agreement with the bulk value 18.80~eV reported on GaAs(110). \cite{Kraut1980}
Comparing $E_{\text{VBM}}$ of GaAs with GaAsBi (2.7\%), a shift to lower BEs is visible in Fig.~\ref{fig:Kraut}, which we estimate to $(E_{\text{VBM}}^{\text{GaAs}} - E_{\text{VBM}}^{\text{GaAsBi}})=(0.23 \pm 0.03)$~eV. 
At the same time, however, $E_{\text{Ga 3d}}^{\text{GaAsBi}}$ values shift in the same direction. Taking into account the limited energy resolution of the laboratory NanoESCA instrument we estimate bulk-related, Bi-induced changes in the relative Ga 3d CL positions $(E_{\text{Ga 3d}} - E_{\text{VBM}})$ to be smaller than 0.1~eV.

Short ion bombardment as described in the above section introduces surface-near defects. Under subsequent annealing cycles at temperatures of 400-440$^\circ$C, surfaces partly reorder in $2\times3$ surface symmetries. Again we want to reference CLs to $E_{\text{VBM}}$, which is shown in Fig.~\ref{fig:Kraut-sputtering} for GaAsBi (2.7\%). During sputtering/annealing the VBM moves towards lower values by $0.3$~eV and Ga 3d and As 3d CLs become broadened, both evidencing an enhanced disorder due to the sputtering procedure. In the case of Bi 5d CLs a strong change of the spectral shape is visible after sputtering and annealing, which corresponds to the suppression of a higher BE component at $25$~eV connected to the metallic phase of Bi as we will show in the following.

%%%%%%%%%%%%%%%%%%%%%%%%%%
\begin{figure}
\includegraphics[width=0.48\textwidth]{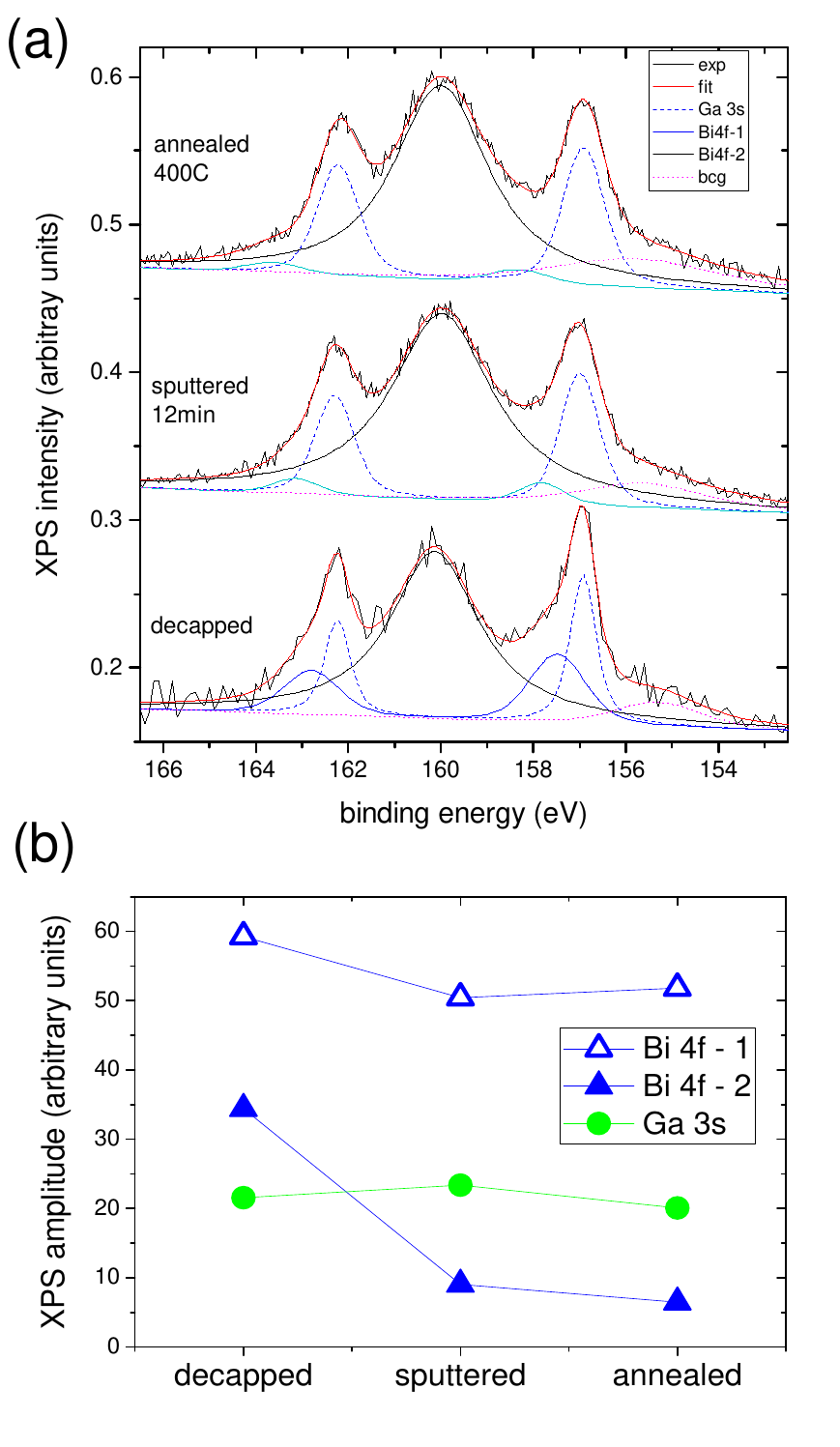}
\caption{(a) Bi 4f XPS spectra of GaAs${_{1-\text{x}}}$Bi$_\text{x}$ (x=2.7\%) after decapping, subsequent 12 min sputtering, and final annealing at 440$^\circ$C (bottom to top). The fit shown to the data consists of one Ga 3s line and two Bi 4f doublets. A broad plasmonic peak is present at about 155~eV, which originates from the As 3$p$ edges at lower BEs. Fit parameters are listed in supplementary information. In (b) the intensities of the 3 fit components are plotted for the three sample stages.}
\label{fig:XPS}
\end{figure}
%%%%%%%%%%%%%%%%%%%%%%%%%%%

%=----------------------------------------------------------------------=
%\subsection{Bismuth 4f states \label{sec:Bi4f}}
%=----------------------------------------------------------------------=
In the initial decapped surface shown in Fig.~\ref{fig:XPS}(a), bottom panel, Bi 4f XPS can be decomposed into two Bi doublets with the main 4f$_{5/2}$ peak located at BEs (157.7 $\pm$ 0.2)~eV and (157.0 $\pm$ 0.1)~eV. The fit was derived using the {\it KolXPD} program and respective parameters are given in supplementary information. The BE position (157.7 $\pm$ 0.2)~eV is close to
literature values attributed to Bi metallic state\cite{Laukkanen2017}. 12 minutes of Argon sputtering almost completely removes the metallic Bi component at (157.7 $\pm$ 0.2) eV (see middle panel), while the Ga 3s reference peak remains unchanged. It proves the surface-near character of the metallic phase. The remaining Bi component at
(157.0 $\pm$ 0.1)~eV reflects the incorporation of single Bi atoms into the host GaAs lattice, most likely Bi\textsubscript{As} substitutional sites, as suggested by Sales et al.\cite{Sales2011}. 
The bulk character of the (157.0 $\pm$ 0.1)~eV component is supported by the fact that no significant differences neither in Bi XPS spectral shapes nor in the ratio between Ga 3s and Bi 4f
intensities $\it r$ = I\textsubscript{Bi4f} / I\textsubscript{Ga3s} appear when sputtering times are varied between 1 min and 12 min (see Fig.~\ref{fig:sputtering-effect-SI} in the supplementary material). Fig.~\ref{fig:XPS}(a)
(top panel) moreover shows that subsequent annealing up to
440$^\circ$C after sputtering cycles does not trigger a reappearance of the
metallic Bi phase in XPS. It proves the absence of major diffusion processes of Bi atoms at temperatures used throughout this work and the presence of negligible amount of Bi clusters in the bulk.

%%%%%%%%%%%%%%%%%%%%%%%%%%%%%%%%%%%%%%%%%%%%%%%%%%%%%%%
\section{Angle-resolved $k$-PEEM studies \label{sec:kPEEM}}
%%%%%%%%%%%%%%%%%%%%%%%%%%%%%%%%%%%%%%%%%%%%%%%%%%%%%%%

%=----------------------------------------------------------------------=
\subsection{GaAs(001) \label{sec:GaAs}}
%=----------------------------------------------------------------------=

%%%%%%%%%%%%%%%%%%%%%%%%%%
\begin{figure*}
\includegraphics[width=0.95\textwidth]{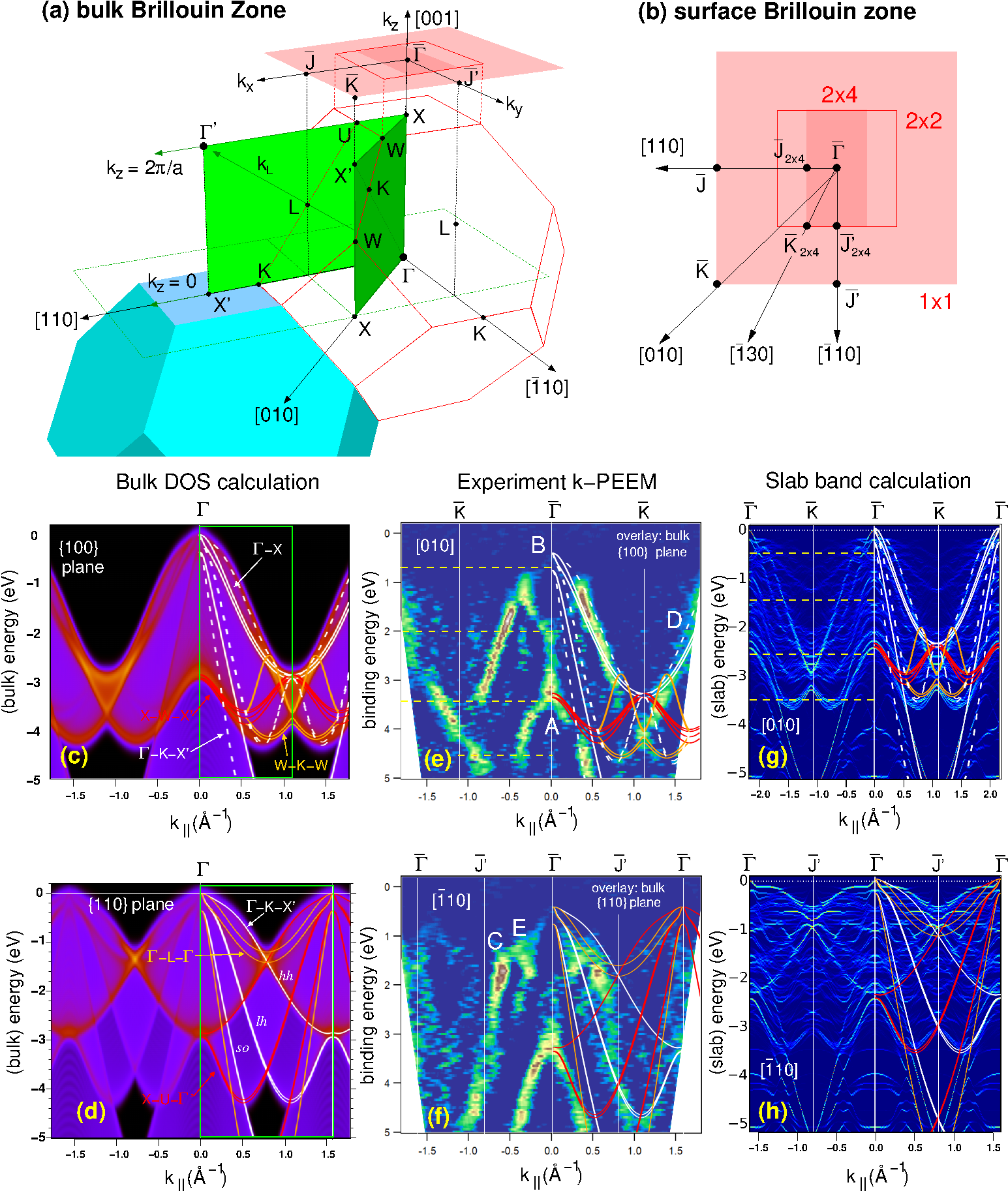} % PDF too large for overleaf. DO NOT CHANGE ASPECT RATIO!
\caption{(a) Bulk and (b) surface Brillouin zones of GaAs(001). 
(c) Bulk GaAs DOS plotted along $[010]$ and averaged over $k_z$, corresponding to the dark-green $\Gamma$XX'X plane shown in (a).
(d) As (c), for the $[110]$ direction and the light-green plane $\Gamma$X'$\Gamma'$X.
Selected bands along high symmetry lines lying in these planes and projected onto $k_{||}$ are overlaid on the right hand side of each panel. 
(e) and (f) Experimental k-PEEM data for [010] and $\bar{1}10]$ directions, with computed bulk bands overlaid as in (c) and (d).
(g) and (h) Unfolded slab band structures for the $\beta_2(2\times4)$ reconstruction for the [010] and [$\bar{1}10]$ directions. Bulk overlays are compressed to match the slab bands.
Horizontal dashed lines in (e) and (g) indicate energies at which ($k_x$, $k_y$) distributions shown in Fig.~\ref{fig:kPEEM-GaAs2} are generated.
Computed energies are shown with respect to the valence band maximum, experimental energies refer to the measured Fermi level. 
}
\label{fig:kPEEM-GaAs1}
\end{figure*}
%%%%%%%%%%%%%%%%%%%%%%%%%

% %%%%%%%%%%%%%%%%%%%%%%%%%%
% \begin{figure*}
% \includegraphics[width=7.2in,height=7.5in]{Images/GaAs-ARPES-theory-overlay+unfolded+DOS.pdf}
% \caption{Extended bulk (a) and surface (b) Brillouin zone of GaAs(001). (c) and (f) shows respective $k$-PEEM data of a $2\times 4$ reconstructed GaAs(001) along [010] and [$\bar{1}$10], which correspond to cuts along the high-symmetry surface Brillouin zone directions $\bar{\Gamma} - \bar{\text{K}}$ and $\bar{\Gamma} - \bar{\text{J}}$ as shown in dark and light green planes in (a). Superimposed are computed bands of bulk GaAs along $\Gamma - \text{X}$ and $\Gamma - \text{K}$ at $k_{\parallel}$ = 0 (white) and $k_{\parallel}$ = 2$\pi/a$ (red). Dashed horizontal lines indicate four energies at which ($k_x$, $k_y$) distributions shown in Fig.~\ref{fig:kPEEM-GaAs2} are generated.
% (b) and (g) show {\it ab initio} calculations along [010] and [$\bar{1}$10] for comparison. Also here bulk calculations are plotted ontop for direct comparison. In (c) and (h) the calculated band structure are shown weighted by the DOS.}
% \label{fig:kPEEM-GaAs1}
% \end{figure*}
% %%%%%%%%%%%%%%%%%%%%%%%%%

 A prerequisite for momentum resolved \emph{k}-PEEM measurements is a sufficient
crystalline surface order of samples under study, which we monitor by
LEED. We first discuss the case of GaAs without Bi. 
After decapping of pure GaAs(001) (0\% Bi) surfaces we observe the well-known ($2\times 4$) reconstruction as shown in Fig.~\ref{fig:manuscript-LEED-image}(a). Its electronic \cite{Cai1992,Chiang1983,Olde1990,Larsen1982} and optical \cite{Placidi2006,Arciprete2004a} properties were long studied during the past decades.
The corresponding ($2\times 4$) SBZ is shown in Fig.~\ref{fig:kPEEM-GaAs1} (a) and (b) relative to that of the $(1\times1)$ SBZ. In the following, special point labels of the ($2\times 4$) SBZ are indicated by the appropriate subscript; otherwise, points refer to the $(1\times1$) SBZ. 

% Conor's suggested text
Experimental \emph{k}-PEEM data for photon energies of $h\nu$ = 21.2~eV are shown in Fig.~\ref{fig:kPEEM-GaAs1} (e) and (f) for $k_{||}$ along the [010] ($\Gbar$--$\Kbar$) and [$\bar{1}$10] ($\Gbar$--$\Jpbar$) directions, respectively. The data cover a range of $k_{||} = \pm$1.8~\AA, centred at $\Gbar$ and BEs are given with respect to the Fermi level of the \emph{k}-PEEM analyser.
Again we estimate the VBM $E_{\text{VBM}}$ using the leading edge method as shown in Fig.~\ref{fig:VBM-kPEEM} based on $k$-space integrated intensities. We derive a value $E_{\text{VBM}}^{\text{GaAs}} = (0.68 \pm 0.02)$~eV, which is close to the value from XPS data although the probing depth of measurements at $h\nu$ = 21.2~eV is significantly lower than in the XPS mode. We infer that band bending effects do not significantly affect values $E_{\text{VBM}}^{\text{GaAs}}$. Typical screening lengths of MBE-grown GaAs are 1-2nm.
%Here some general description of the spectra

%and thus probe (at least) the full ($1\times1$) SBZ.
%for the [$\bar{1}$10] case in Fig.~\ref{fig:kPEEM-GaAs1}(f) the 2\textsuperscript{nd} zone $\Gbar$ points are visible thus confirming the $k$-space calibration of the NanoESCA electron lens system. 
Both cuts show similar band structures with several common features that are marked \textit{A}-\textit{E} (partly following the notation of Souma et al.~\cite{Souma2016}), although the [010] cut appears somewhat narrower at lower BEs. Band {\it B} shows a maximum at
$\Gamma$ and continuously disperses down to BE = 3.5~eV. A characteristic `wishbone' feature \textit{A} is observed below 3.2~eV at $\Gbar$.
%(indicated in Fig.~\ref{fig:kPEEM-GaAs2} by a dashed line). 
%It is observed in all cuts including the bulk high
%symmetry direction $\Gamma$-K\textsubscript{$1\times 1$}.

At these incident energies we expect a dominant contribution from bulk states, and thus associate the maximum at $\Gbar$ with the peak in the light-hole or split-off band of bulk GaAs. To understand the data further, one must consider the relation between the $(1\times1)$ SBZ and the bulk GaAs BZ as depicted in Fig.~\ref{fig:kPEEM-GaAs1} (a). 
It is well known that in \emph{k}-PEEM, as in ARPES, the momentum $k_z$ perpendicular to the surface is typically not conserved. In order to interpret the measured data along some $k_{||}$, one should thus in principle consider a suitable average over $k_z$.\cite{Kumigashira1998} For instance, a measurement along [010] ($\Gbar$--$\Kbar$) will probe all $(k_{||},k_z)$ in the $\Gamma$XX'X plane (see dark-green plane in Fig.~\ref{fig:kPEEM-GaAs1} (a)). 

As a first approximation of the \emph{k}-PEEM data, we computed within DFT (including spin-orbit coupling) the density of states of bulk GaAs resolved on a $(100 \times 40)$ grid of $k_{||},k_z$ points along the [010] and [110] directions, before taking a simple average over $k_z$ for each value of $k_{||}$. The results are plotted as colormaps in Fig.~\ref{fig:kPEEM-GaAs1}(c) and (d), and extended across the measured range of $k_{||}$. Overlaid on these maps are the computed bulk bands along selected high-symmetry lines. It appears that most of the $k_{||}$-resolved DOS can be well represented by a handful of selected bands passing through the high-symmetry points of the bulk BZ.\cite{Kumigashira1998,Souma2016}.

In order to facilitate a better comparison with experiment, these selected bands are also overlaid on the \textit{k}-PEEM data in Fig.~\ref{fig:kPEEM-GaAs1}(e) and (f). The $k_{||}$-resolved DOS clearly succeeds in explaining most of the observed signal. 
Feature \textit{A} at $\Gbar$ arises from the contributions at $k_z=2\pi/a$, i.e. the lines X-W-X' and X-U-$\Gamma$' drawn in red. We stress that \textit{A} can only be explained by considering the $k_z$ dependence.\cite{Souma2016} The intense signal spreading out symmetrically below \textit{A} is explained first by the dispersion of the $k_z=2\pi/a$ lines and then, for higher binding energies, by the diagonal projections $\Gamma$-K-X' [in (c)] and $\Gamma$-L-$\Gamma$ [in (d)]. The intense vertical line above \textit{A} (BE = 2.0--3.4~eV) in the [010] case instead cannot be associated with a single band, but corresponds well to a region of high DOS visible only in the $k_{||}$-resolved map (c).
Features \textit{C} and \textit{D} are well explained by heavy-hole or light-hole bands arising from the $k_z=0$ lines. Note that the horizontal signal `widths' for both directions are determined by the $\Gamma$-K-X' line. However, its contribution is projected onto $k_{||}$ for the [010] case, resulting in the overall narrower lineshape observed experimentally.

Good agreement with experiment is also obtained for the DFT band structure of the full GaAs(001)-$\beta2(2\times4)$ slab, plotted in  Fig.~\ref{fig:kPEEM-GaAs1}(g) and (h) and unfolded across the full $(1\times1)$ BZ. Due to the use of a relatively thin slab, any extended (bulk-like) wavefunctions undergo size quantization and thus the computed band structure is compressed relative to the bulk GaAs case. This is illustrated by overlaying and compressing the selected bulk bands until a reasonable match is reached (see Figure). Note that the slab bands naturally include the $k_z$ averaging procedure as the 3D BZ of the large (slab plus vacuum) supercell is significantly compressed, eventually yielding a good approximation to the SBZ itself. In addition to the dispersive bulk bands, several reconstruction-related features become apparent. A dispersive band with minimum around 1.1eV at $\Kbar$ is visible along [010], while flatter features appear in the 0.8--1.2eV range along [$\bar{1}$10]. The latter in particular is suggestive of surface states.
%, and will be discussed further in Section.~\ref{sec:UPS}.

%%%%%%%%%%%%%%%%%%%%%%%%%%%
\begin{figure}
\includegraphics[width=3.5in,height=2in]{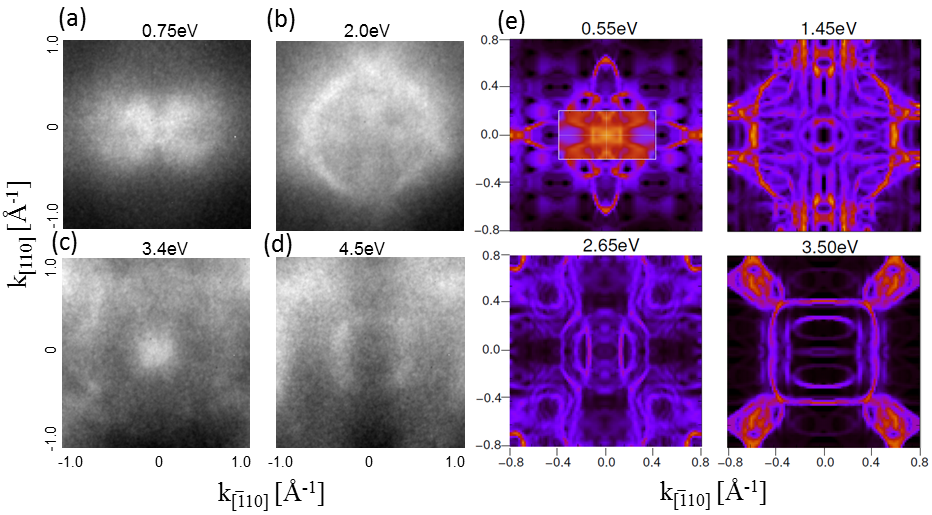}
\caption{Left: Constant energy cuts of  pristine GaAs(001) k-PEEM data. Intensity distributions (k\textsubscript{x}, k\textsubscript{y}) are shown at four BEs (a) 0.75~eV, (b) 2.0~eV, (c) 3.4~eV, and (d) 4.5~eV. 
Right: band structure maps of the GaAs(001)-$\beta_2(2\times4)$ slab, computed at constant energies indicated in Fig.~\ref{fig:kPEEM-GaAs1}(g), and unfolded across the $(1\times1)$ SBZ. 
Panel (e) inset shows the (k\textsubscript{x}, k\textsubscript{y})-resolved DOS of the $(2\times4)$ SBZ. Theoretical energies in (e) are referenced to the VBM, while experimental data (a)-(d) are given with respect to the Fermi level.
\label{fig:kPEEM-GaAs2}
}
\end{figure}
%%%%%%%%%%%%%%%%%%%%%%%%%%%%%

% In the $2\times 4$ reconstruction the bulk symmetry between [110] and [$\bar{1}$10]
% directions is broken and respective surface electronic properties should
% differ.
% %% Here we should cite ourselves PRB_69_081308(R)_(2004)
% In Fig ~\ref{fig:kPEEM-GaAs1}(c)-(e) high-symmetry (\emph{E}, \emph{k}) cuts along the
% $\Gamma$-J'\textsubscript{$1\times 1$}, $\Gamma$-J\textsubscript{$1\times 1$}, $\Gamma$-\textsubscript{$2\times 1$}-K directions are shown, respectively,  which were derived from 3-dimensional (E, k\textsubscript{x}, k\textsubscript{y}) k-PEEM data
% taken at photon energies  $h\nu$ = 21.2~eV. For comparison the cuts along the
% bulk high symmetry direction $\Gamma$-K\textsubscript{$1\times 1$} is shown in (f).

% Other directions
In the $2\times 4$ reconstruction the bulk symmetry between [110] and [$\bar{1}$10] directions is broken and respective surface electronic properties should differ.
Figure \ref{fig:ARPES-GaAs-GaBiAs-comp} adds \emph{k}-PEEM data also for the [110] and [$\bar{1}$30] directions. % Is [100] a mistake? No, it's -130. Hello! No, I mean, the [100] in panel (a) looks identical to the [010] in Fig 5. Hi Conor, yes it's a mistake, the right notation in panel (a) fig. 6 is [010]!
Note that signals from $\Gbar$ in the 2\textsuperscript{nd} SBZ are visible in both (c) [110] and (d) [$\bar{1}$10] data thus confirming the $k$-space calibration of the NanoESCA electron lens system. 
% Cuts along $\Gamma$-J`\textsubscript{$1\times 1$} and
% $\Gamma$-J\textsubscript{$1\times 1$} directions in Fig.~\ref{fig:kPEEM-GaAs1} (c) and (d) clearly show the 2\textsuperscript{nd} gamma points $\Gamma$', which confirm the $k$-space calibration of the NanoESCA's electron lens system. 
%
% What about showig intead the [110] in the first figure, since "E" is most clear in the -110 direction?
Both cuts show very similar band structures, consistent with our explanations of the main features as arising from the bulk DOS averaged in the \{110\} plane. The `wishbone' feature in the [110] data is now more clearly shown to be derived from two distinct bands that cross at around 3.7~eV. 
% with three distinct band features, which we label as {\it A}, {\it B},
% and {\it C}, following the notation of S. Souma et al.~\cite{Souma2016}. Band {\it B} shows a maximum at
% $\Gamma$ and continuously disperses down to EB = 3.0~eV (indicated in Fig.~\ref{fig:kPEEM-GaAs2} by a dashed line). 
Band \textit{B} is observed in all four cuts.
% including the bulk high symmetry direction $\Gamma$-K\textsubscript{$1\times 1$}.
%Its effective mass of about
%????? m0 (m0: the free-electron mass) is in good agreement with the one of
%the split-off (SO) band\cite{Souma2016}. 

Further confirmation of the origin of these bands is obtained by performing constant energy cuts of the k-PEEM data.
Exemplary energy cuts at BE = 0.75~eV, 2.0~eV, 3.4~eV and 4.5~eV
are shown in Fig.~\ref{fig:kPEEM-GaAs2}(a)-(d), and confirm asymmetries between the [110] and
{[}$\bar{1}$10{]} directions. The measurements are in generally good agreement with appropriate energy cuts through the unfolded slab band structures (see Fig.~\ref{fig:kPEEM-GaAs1} (g) and (h)). 
It is worth noting that in the energy cut at BE = 0.75~eV in Fig.~\ref{fig:kPEEM-GaAs2} the (k\textsubscript{x},
k\textsubscript{y}) intensity
reveals a cloverleaf structure which is elongated along the {[}$\bar{1}$10{]} direction
with respect to {[}110{]}. 
This pattern is also evident in the $(k_x,k_y)$-resolved DOS at 0.55~eV of the (folded) $(2\times4)$ slab, shown as an inset in Fig.~\ref{fig:kPEEM-GaAs2}(e).
%[Needs explicit label].
% where is the inset?? Conor: the first theory panel, 2x4 box. Fig 7 needs (e)-(h) panel labels for clarity.
The bulk GaAs electronic structure close to
VBM is known to have As 4\emph{p} character with four-fold cubic symmetry
in the bulk, while for GaAs $2\times 4$ surfaces we expect such a symmetry breaking. 
Band \textit{C} is assigned to the light-hole bands supported by
the rather spherical shape in the ($k$\textsubscript{x}, $k$\textsubscript{y}) distribution at BE$ = 2.0$~eV (see cut shown in Fig.~\ref{fig:kPEEM-GaAs2}(b)). 
Heavy hole bands on the other hand were shown to become visible only at higher photon energies as shown \textit{e.g} by Kanski et al. \cite{Kanski2017}. 

In addition to the above bulk-derived features, we also identify an additional feature \textit{E} in Fig.~\ref{fig:kPEEM-GaAs2} which crosses \textit{B} at BE$ = 1.8$~eV and $k=\pm 0.4$ ~\AA$^{-1}$, reminiscent to a backfolding effect to the 1st BZ. This feature is missing from the bulk DOS calculations, and has not been reported in previous literature\cite{Souma2016}. It may be consistent with the aforementioned feature at $-1$eV in the slab band calculations (Fig.~\ref{fig:kPEEM-GaAs1}(g)) that are suggestive of surface state origin.
\subsection{Bi-doped GaAs(001) \label{sec:GaAsBi}}
%=----------------------------------------------------------------------=

\begin{figure*}
\includegraphics[width=7.2in,height=4in]{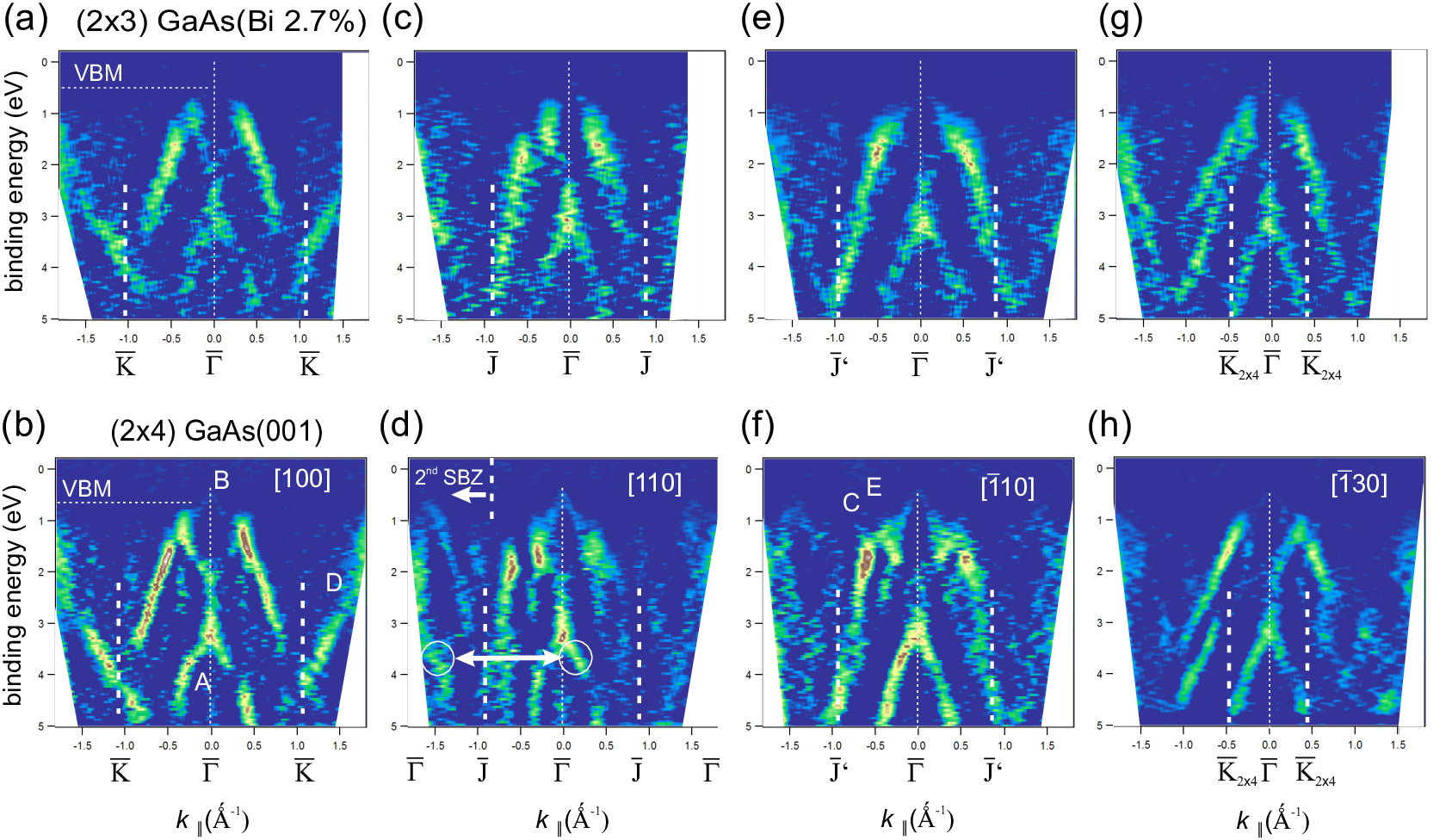}
\caption{$k$-PEEM comparison of pristine ($2\times 4$) GaAs(001) (left column) and sputter-annealed ($1\times 3$) GaAsBi (right column). 
(a)-(b) correspond to cuts along the $\Gbar$--$\Kbar$, 
(c)-(d) $\Gbar$--$\Jbar$, 
(e)-(f) $\Gbar$--$\Jpbar$, and 
(g)-(h) $\Gbar$--$\Kbar_{2\times 4}$ directions. 
Horizontal dashed lines correspond to VBM values $E_{\text{VBM}}$ derived from Fig.~\ref{fig:VBM-kPEEM}. The horizontal arrow in (c) indicates the replica of feature A in the 2nd SBZ at the expected distance $\Delta k=1.56$\AA$^{-1}$.
%\textbf{CH: I believe (b) should read [010], not [100]. Also: in (d) why don't the vertical white lines coincide? (2nd SBZ, Jbar)}
\label{fig:ARPES-GaAs-GaBiAs-comp}
}
\end{figure*}

We now turn to the effect of Bi doping on the GaAs band structure. Fig.~\ref{fig:ARPES-GaAs-GaBiAs-comp} (top row) summarizes \textit{k}-PEEM data of {\it sputter-annealed} GaAs${_{1-\text{x}}}$Bi$_\text{x}$ (x=2.7\%) in comparison with {\it decapped} (unsputtered) GaAs(0\% Bi) (bottom row) for high-symmetry directions $\Gbar$--$\Kbar$, $\Gbar$--$\Jbar$, $\Gbar$--$\Jpbar$, and $\Gbar$--$\Kbar_{2\times 4}$. Respective $2\times 3$ and $2\times 4$ LEED symmetries shown in Fig.\ref{fig:manuscript-LEED-image}(a) and (b) were discussed earlier.

\begin{figure}
\includegraphics[width=2.7in,height=1.9in]{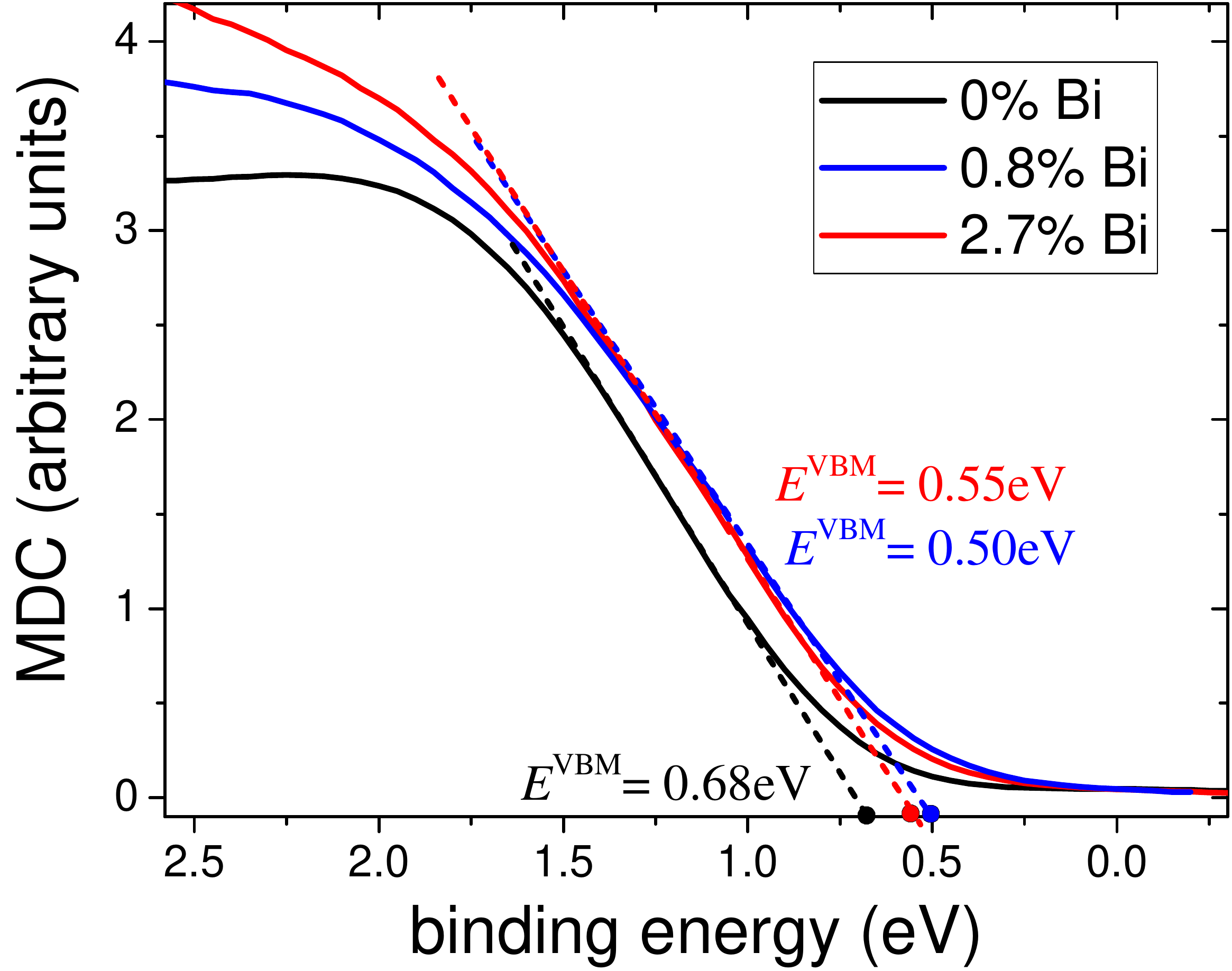}
\caption{Momentum integrated $k$-space intensity close to the VBM plotted vs energy. The photon energy is $h\nu$ = 21.2~eV. Dotted lines show the estimation of the VBM according to the leading edge method (Kraut method).
\label{fig:VBM-kPEEM}
}
\end{figure}

Rigid shifts of the $k$-PEEM band structures of GaAsBi to lower BEs with respect to pure GaAs confirms our XPS observations. Fig.~\ref{fig:VBM-kPEEM} shows a reduced value of $E_{\text{VBM}}^{\text{GaAsBi}} = (0.55 \pm 0.02)$~eV. If we assume that band bending is negligible, the shift of the VBM to lower BEs suggests the formation of shallow defects states, which would pin $E_F$  closer to the VBM.

%%%%%%%%%%%%%%%%%%%%%%%%%%
%%%%%%%%%%%%%%%%%%%%%%%%%

%CH: Here need to discuss all the visible and measured differences. Blurring, shifts, weaker intensity...
The shift in the VBM is accompanied by a general broadening of $k$-PEEM features compared to pure GaAs despite the well-defined LEED pattern in Fig.\ref{fig:manuscript-LEED-image}(b). In particular along the $\Gbar$--$\Kbar$ and $\Gbar$--$\Kbar_{2\times 4}$ directions the band structure of GaAsBi appears to match that of GaAs but considerably broadened. Looking closer, the feature $E$ along the $\Gbar$--$\Jpbar$ direction is particularly affected [see Fig.~\ref{fig:ARPES-GaAs-GaBiAs-comp}(f)], which seems to have entirely disappeared. It could point towards surface related broadening of the feature $E$. It would confirm the above discussed interpretation of our ${2\times 4}$ GaAs slab calculations in the direction of a surface state origin of $E$.

%%%%%%%%%%%%%%%%%%%%%%%%%%
\begin{figure}[tb!]
\includegraphics[width=0.49\textwidth]{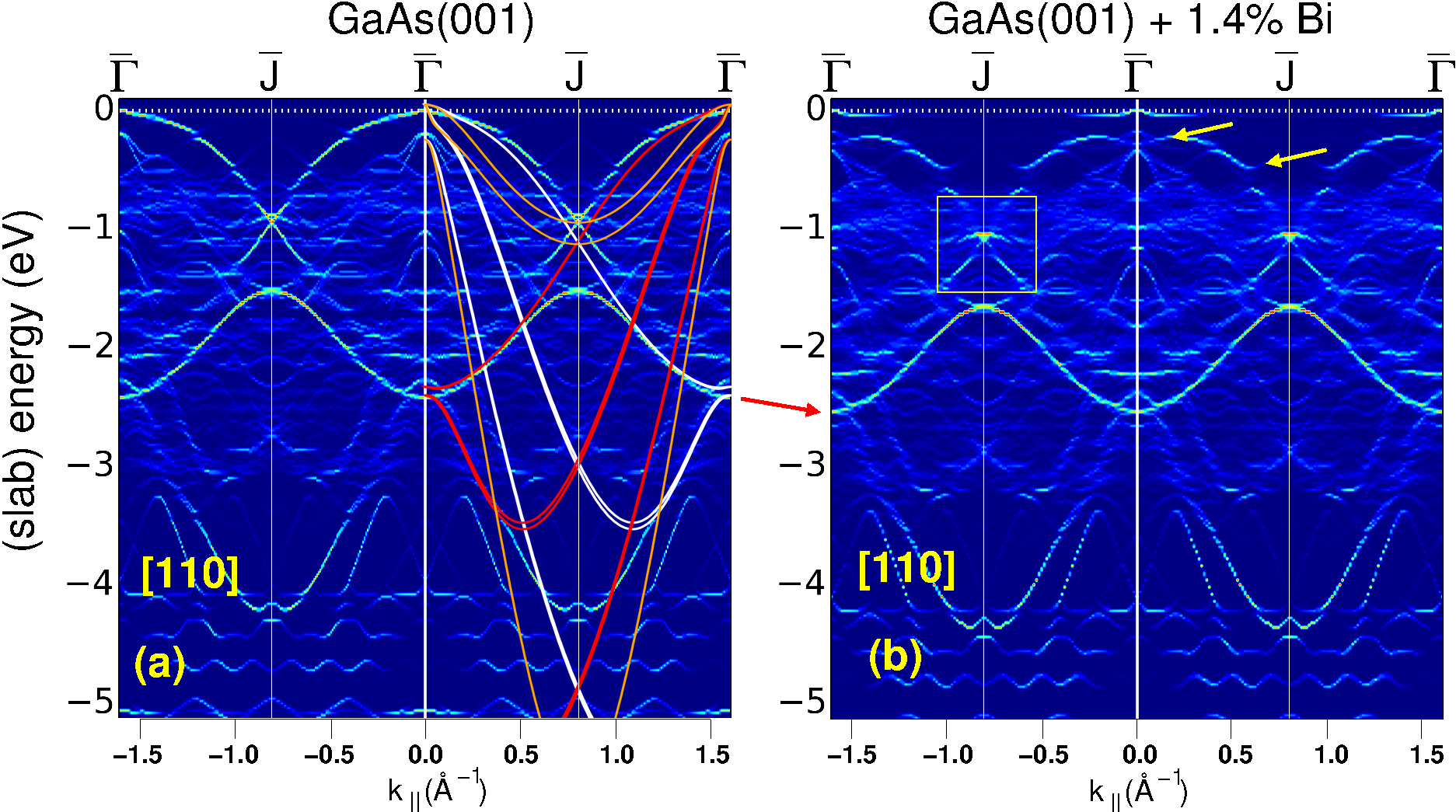}
\caption{
Band structure of GaAs(001)-$\beta2(2\times4)$ slab unfolded along [110]: (a) pristine $(2\times4)$ cell and (b) including 1 atom of Bi in a $(4\times4)$ supercell.
\label{fig:unfolded-GaAsBi}
}
\end{figure}
%%%%%%%%%%%%%%%%%%%%%%%%%

Slab calculations incorporating Bi in a bulk site succeed in reproducing the main observed experimental features. Figure.~\ref{fig:unfolded-GaAsBi} compares the pristine $\beta2(2\times4)$ slab band structure unfolded along [110] with that of a slab containing a low concentration of Bi. The main effect of Bi is to perturb the bands coming from bulk GaAs, in particular the $hh$ band (yellow arrows). Other points in the SBZ also lose their clear Bloch character (box). A small shift or change in the bandwidth is also observed (red arrow). However, due to the dense manifold of bulk and surface bands in these supercells it is difficult to make any quantitative analyses: this is discussed instead in the following section.

%%%%%%%%%%%%%%%%%%%%%%%%%%%%%%%%%%%%%%%%%%%%%%%%%%%%
\section{Ultraviolet photoelectron spectroscopy studies \label{sec:UPS}}
%%%%%%%%%%%%%%%%%%%%%%%%%%%%%%%%%%%%%%%%%%%%%%%%%%%%

%%%%%%%%%%%%%%%%%%%%%%%%%%%
%%%%%%%%%%%%%%%%%%%%%%%%%%%%%

The $k$-PEEM measurements reported in the previous section succeeds in identifying mainly bulk-derived electronic states in GaAs and GaAsBi. However, the limited resolution makes it difficult to identify surface related features previously identified, for instance, with ARPES\cite{Larsen1982}. In order to examine the surface electronic structure in more detail we measure UPS in a highly-resolved mode at the ELETTRA synchrotron facilities. In particular the possibility to vary the photon energy at the synchrotron allows to probe the dimensionality of surface states. True surface states do not disperse with photon energy due to their strict two-dimensionality. On the other hand for surface resonances a strong hybridization with the bulk band structure destroys the two-dimensionality and dispersive effects can be expected. 
% CH: Or something like that!

Fig.~\ref{fig:UPS-GaAsBi} shows UPS data in normal emission taken at selected photon energies between 25~eV and 100~eV. Bi containing samples were measured in the {\it sputter-annealed} state. At high binding energies $> 3$~eV the spectra of all samples are dominated by two weakly dispersive features at 7.6~eV and 12.3~eV (labeled $d$ and $e$) and one strongly dispersing band $c$ which joins $d$ at a photon energy of 40~eV. In addition pure GaAs shows a prominent feature $b$ at 4.2~eV, which is strongly suppressed for 0.8\% Bi and fully disappears at 2.7\%.

% Dispersion $k_{\perp} (h\nu)$ in normal incidence photoemission can be described as:
% \begin{equation}
% k_{\perp}  = \sqrt{(2m/\hbar^2)(E_{\text kin}+\vert V_{\text 0}\vert-g_{\vert\vert}^2}-G_{\perp} \label{eq:kperp}
% \end{equation}
% where $k_{\perp} $ is the wave-vector component of electrons perpendicular
% to the surface, $E_{\text kin}$ the kinetic energy, $V_{\text 0}$ denotes the inner potential referred to the vacuum level, and $G_{\perp}$ is
% the normal component of a bulk reciprocal-lattice vector
% $G = G_{\perp}+G_{\parallel}$. $g_{\parallel}$ is in general a linear combination of the parallel component of the reciprocal bulk lattice vector $G$ and a reciprocal-lattice vector of the ideal or the reconstructed surface. 

%%%%%%%%%%%%%%%%%%%%%%%%%%%
\begin{figure}
\includegraphics[width=2.7in,height=1.9in]{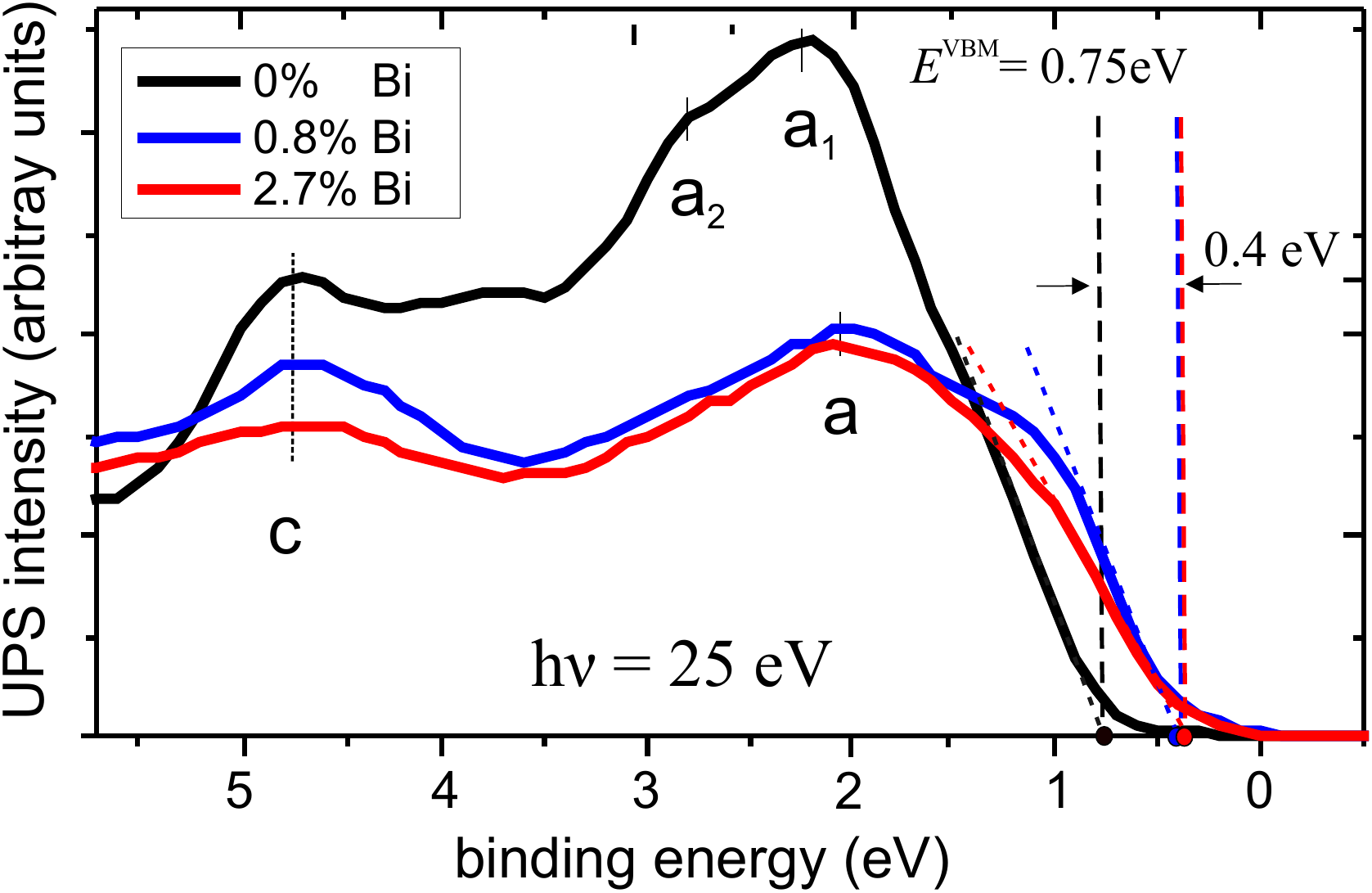}
\caption{Normal emission UPS close to the VBM for Bi concentrations 0\%, 0.8\% and 2.7\% . The photon energy is $h\nu$ = 25~eV. Values for $E_{\text{VBM}}$ were derived using the leading edge method (Kraut method) as shown by vertical lines in the plot.}
\label{fig:UPS-VBM}
\end{figure}
%%%%%%%%%%%%%%%%%%%%%%%%%%%%%

Spectra taken from {\it decapped} GaAs(001) are in good agreement with the literature \textit{e.g.} for GaAs(001)-$2\times 4$ at $h\nu$ = 29~eV ~\cite{Larsen1982}. According to the literature the bands $c$ and $d$ are primary cone bulk emission peaks and can be attributed to direct interband transitions from bulk valence bands to a free-electron like final state. The respective dispersive behavior of the bands versus photon energy is plotted in Fig.~\ref{fig:UPS-GaAsBi-dispersion}(a). We have referenced the band dispersion to respective VBM values $E_{\text{VBM}}^{\text{GaAs}} = 0.75$~eV derived in Fig.~\ref{fig:UPS-VBM} using the leading edge method. $c$ shows strong dispersion and a shallow minimum at 43~eV, while $d$ is flat. 

Both findings coincide very well with Cai et al ~\cite{Cai1992} on GaAs(001)-$1\times 1$, confirming that $a$ and $d$ are independent of the particular surface reconstruction as expected for bulk bands. The minimum of $c$ at a photon energy of about 44~eV corresponds to the $X_6$ point in $k_{\perp}$ along the $\Gamma\Delta X$ direction (see label in Fig.~\ref{fig:UPS-GaAsBi-dispersion}(a)) and is attributed to the $\Delta_1$ band. 
Close to the VBM the UPS data of pure GaAs show two features $a_1$ and $a_2$, which join at low photon energies $\le 23$~eV to form one single band $a$ (see label in Fig.~\ref{fig:UPS-GaAsBi-dispersion}(a)). Again this is in good agreement with Cai et al. and suggests that $a_2$ corresponds most likely to the $\Delta_{3,4}$ bulk band transition.
%%%%%%%%%%%%%%%%%%%%%%%%%%%
\begin{figure*}
\includegraphics[width=7in,height=3.3in]{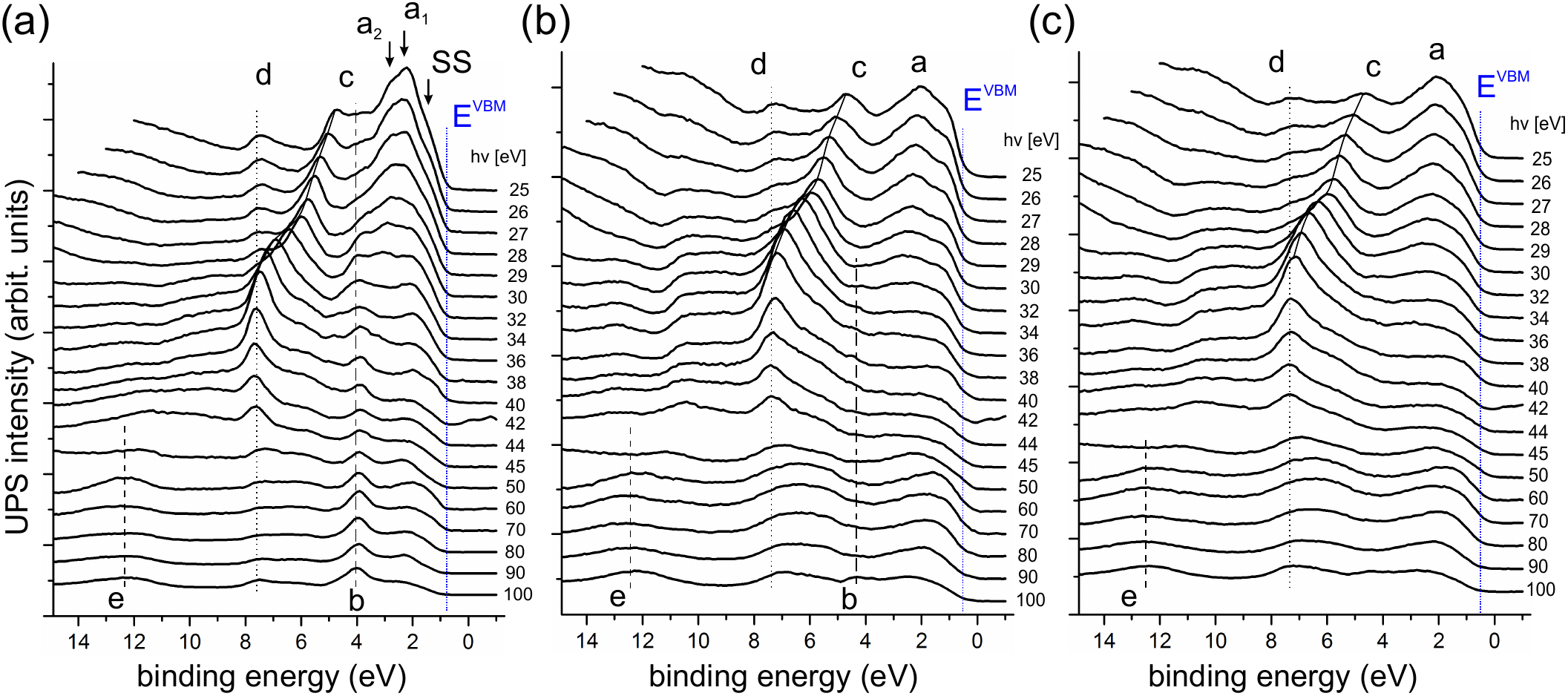}
\caption{Photon energy dependent normal emission UPS spectra for 0\%, 0.8\% and 2.7\% Bi concentrations (a)-(c), respectively. Bi-doped samples are in the {\it sputter-annealed} state. Indications $a$-$e$ and are explained in the text. Vertical dashed lines correspond to VBM values $E_{\text{VBM}}$ derived from Fig.~\ref{fig:UPS-VBM}.}
\label{fig:UPS-GaAsBi}
\end{figure*}
%%%%%%%%%%%%%%%%%%%%%%%%%%%%%
%%%%%%%%%%%%%%%%%%%%%%%%%%%
\begin{figure}[tb!]
\includegraphics[width=2.8in,height=3.5in]{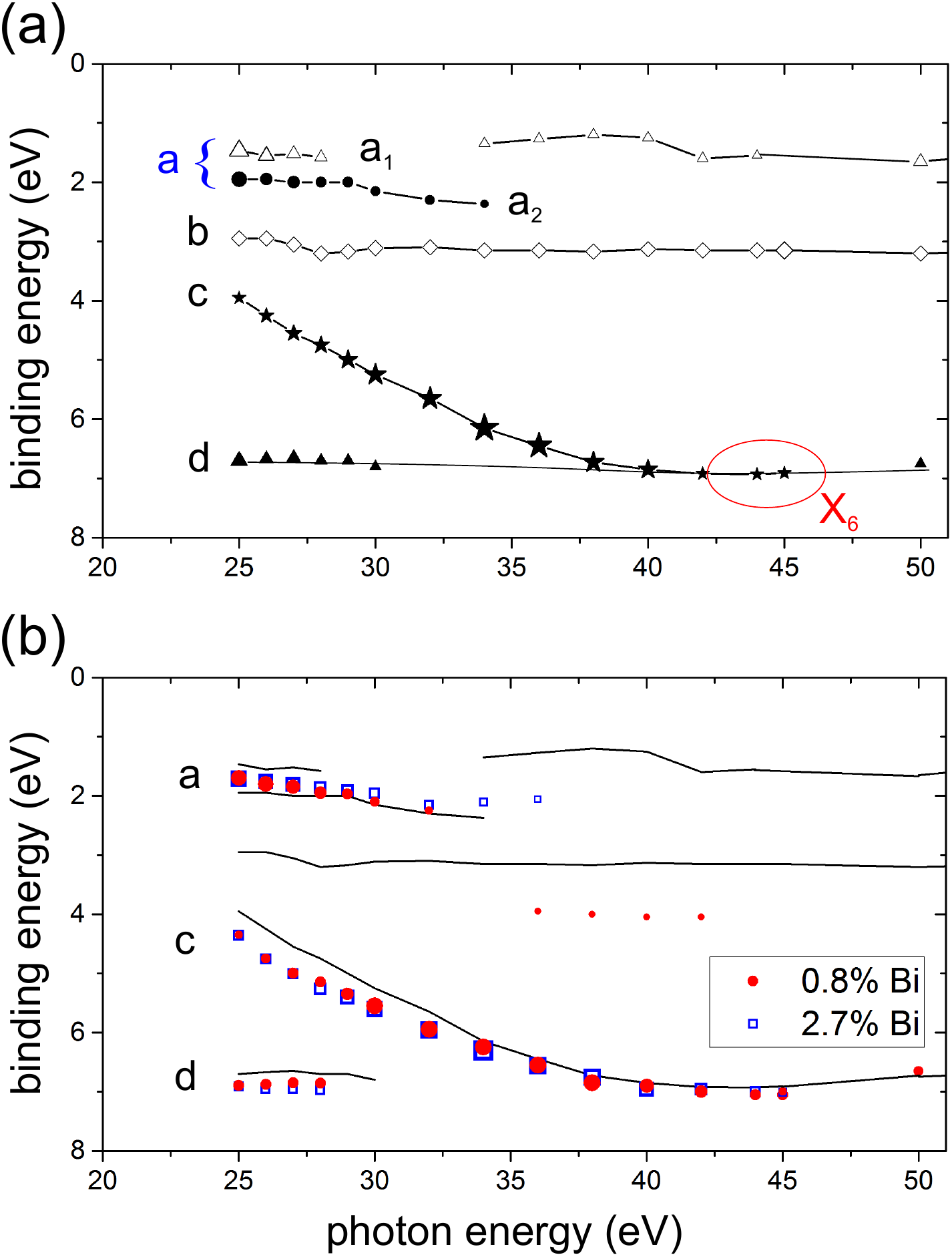}
\caption{Dispersion of bands derived from normal emission UPS data in Fig.~\ref{fig:UPS-GaAsBi}. Symbol sizes reflect the intensity of the band signal at respective photon energies. (a) shows GaAs(001)-2x4 and (b) a comparison with Bi doped samples. In (b) the data of (a) is included as lines. All BEs in this plot are given with respect to VBM values $E_{\text{VBM}}$ in Table~\ref{tbl:VBMvalues}.}
\label{fig:UPS-GaAsBi-dispersion}
\end{figure}
%%%%%%%%%%%%%%%%%%%%%%%%%%%%%

Fig.~\ref{fig:UPS-GaAsBi-dispersion}(b) shows the band dispersions of Bi-doped samples with 0.8\% and 2.7\% Bi concentrations. Data of pure GaAs from (a) is replotted in as lines for direct comparison. 
Comparing the two Bi-doped samples it is striking that the two data sets mostly coincide. Also respective VBM values for Bi-doped samples are similar with $E_{\text{VBM}}^{\text{GaAs}_{0.992} \text{Bi}_{0.008}} = 0.35$~eV and $E_{\text{VBM}}^{\text{GaAs}_{0.973} \text{Bi}_{0.027}} = 0.34$~eV shown in Fig.~\ref{fig:UPS-VBM}. 
In comparison with pure GaAs(001)-$2\times 4$, however, significant differences are visible. ($a_1$ , $a_2$) bands closer to the VBM are broadened and cannot be distinguished anymore as shown on a larger scale in Fig.~\ref{fig:UPS-VBM}. We therefore label them henceforth as $a$. The dispersion of $a$ comes to lie between $a_1$ and $a_2$ and suggests minor shifts in the bulk band $\Delta_{3,4}$ under the influence of Bi although the general broadening limits the certainty of this statement.
On the other hand bands $c$ and $d$ are clearly shifted to higher BEs for Bi containing samples. 
%Nevertheless, the data points lay about 0.4~eV above $a_2$ for pure GaAs, and thus our data would not be against a shift of $\Delta_{3,4}$ in the same direction of $\Delta_{1}$ when GaAs is doped with Bi.
Particularly large is the shift of $c$ of about +0.4~eV, while $d$ shifts by only +0.2~eV.

So far we have neglected surface states (SSs), which contribute to the density of states close to the VBM and should be most affected by reconstruction effects. It is known from the literature on various GaAs surfaces that SSs become visible as typical shoulders~\cite{Cai1992, Olde1990} in the UPS intensity close to the VBM. In Fig.~\ref{fig:UPS-GaAsBi} such a shoulder is labeled SS as an example which is clearly separated from the intensities $a_1$ and $a_2$ at low photon energies. The introduction of Bi in the GaAs matrix affects the VB lineshape, an this distinction in Fig.~\ref{fig:UPS-VBM} is less pronounced.
Upon Bi doping the SS shoulder partly overlaps with the broadened contribution $a$ and it is shifted to lower BEs. In contrast, intensities $c$ representing bulk contributions $\Delta_1$ remain at the same BE as discussed earlier. 
%Continue ... Discussion ... 4kBT ~ 0.1eV broadening....  

%%%%%%%%%%%%%%%%%%%%%%%%%%%
\begin{figure}[tb!]
\includegraphics[width=8.6cm]{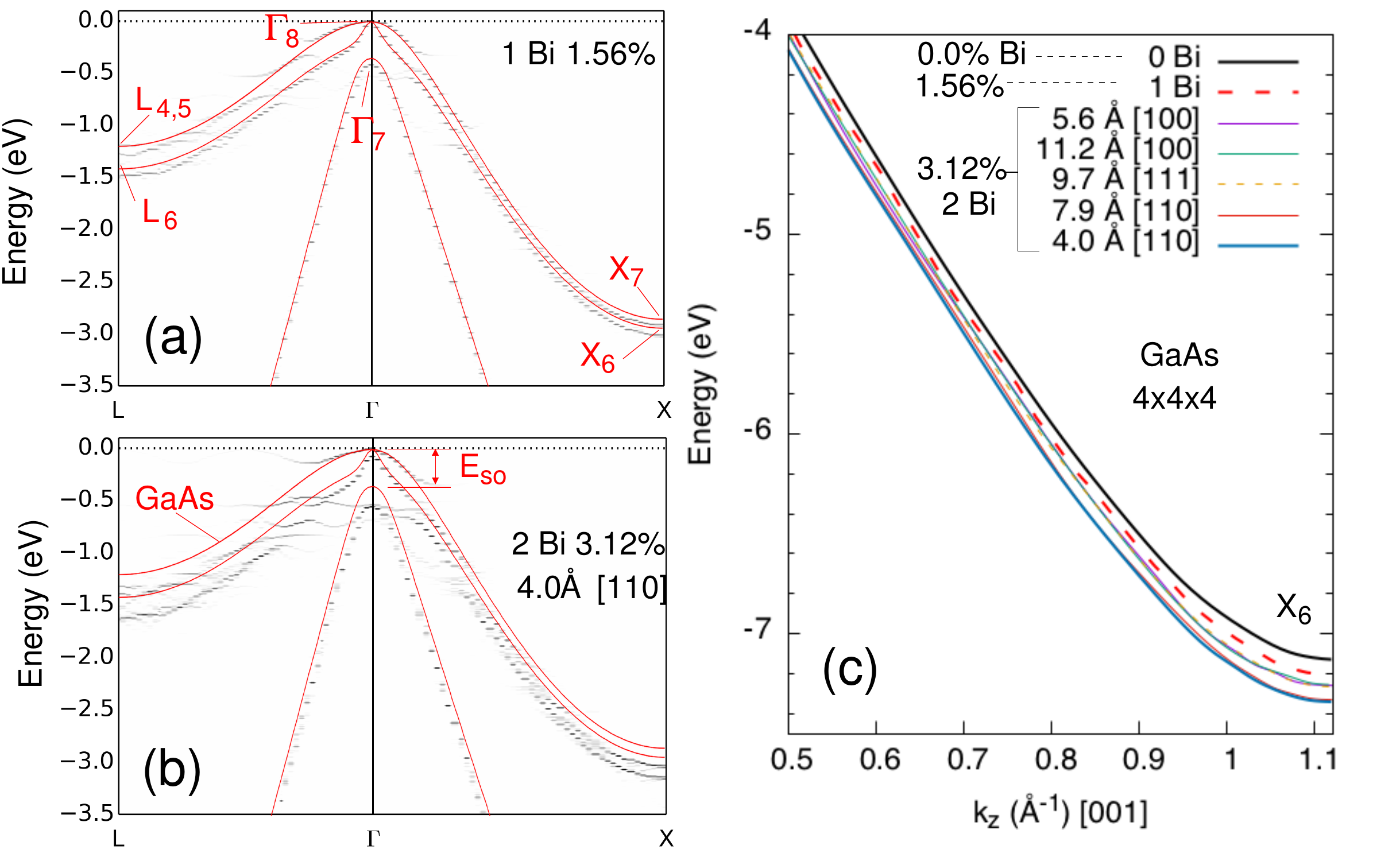}
\caption{Unfolded band structures of GaAs and GaAsBi computed in a 4x4x4 bulk supercell, for (a) 1 Bi atom per cell and (b) 2 Bi atoms per cell (bands of pristine GaAs are overlaid in red). 
(c) Split-off band along $\Gamma-X$ for increasing Bi composition. Several Bi--Bi orientations and distances are indicated. 
}
\label{fig:GaAsBi-4x4x4}
\end{figure}
%%%%%%%%%%%%%%%%%%%%%%%%%%%%%
Although incorporation of Bi in the GaAs(001) slab calculations does yield tangible changes (see Fig.~\ref{fig:unfolded-GaAsBi}), the strong folding makes it difficult to determine the precise influence of Bi-incorporation on specific bulk bands.
We thus performed supercell calculations of \emph{bulk} GaAs and performed a basic analysis of Bi doping on the bulk electronic properties, similar to the study of Bannow et al~\cite{Bannow2016}.
Fig.~\ref{fig:GaAsBi-4x4x4} demonstrates the influence of Bi doping for 1 and 2 Bi atoms per 128-atom $4\times4\times4$ supercell, consistent with concentrations of 1.56\% and 3.12\% Bi, respectively.

Panels (a) and (b) demonstrates, for increasing Bi concentrations (i) significant Bi-induced perturbation on the lh and hh bands near the VBM; (ii) increasing split-off energy $\Delta_\mathrm{SO}$; (iii) away from $\Gamma$, shifts of all bands to higher binding energies; (iv) the SO band is relatively unperturbed. These observations are consistent with the unfolded slab bands in Fig.~\ref{fig:unfolded-GaAsBi}. Regarding (iii), we find a relative shift of the X$_6$,$X_7$ points of about 0.1--0.2 eV at 3.12\% Bi. Closer to $\Gamma$, however, the unfolded bands split yielding features at lower binding energies. This is somewhat consistent with the observed behaviour of a$_2$ at low photon energies. Fig.~\ref{fig:GaAsBi-4x4x4}(c) instead shows the behaviour of the $\Gamma$--X split-off band at higher photon energies, for various Bi--Bi configurations. The directional dependence of the Bi--Bi defect close to $\Gamma$ has previously been noted.\cite{Bannow2016} We confirm this dependence in our data at the X$_6$ point around $-7$eV, with largest shifts occurring for Bi--Bi oriented along [110]. The magnitude of the shift, 0.1--0.2 eV, is comparable to that observed experimentally in Fig.~\ref{fig:UPS-GaAsBi-dispersion}(b), which obviously refers to a quasi-random Bi distribution at an average concentration of 2.7\% Bi. Our bulk supercell calculations therefore succeed in reproducing the basic Bi-induced deformations of the GaAs bands detected in UPS.

%%%%%%%%%%%%%%%%%%%%%%%%%%%
\begin{figure*}
\includegraphics[width=0.8\textwidth]{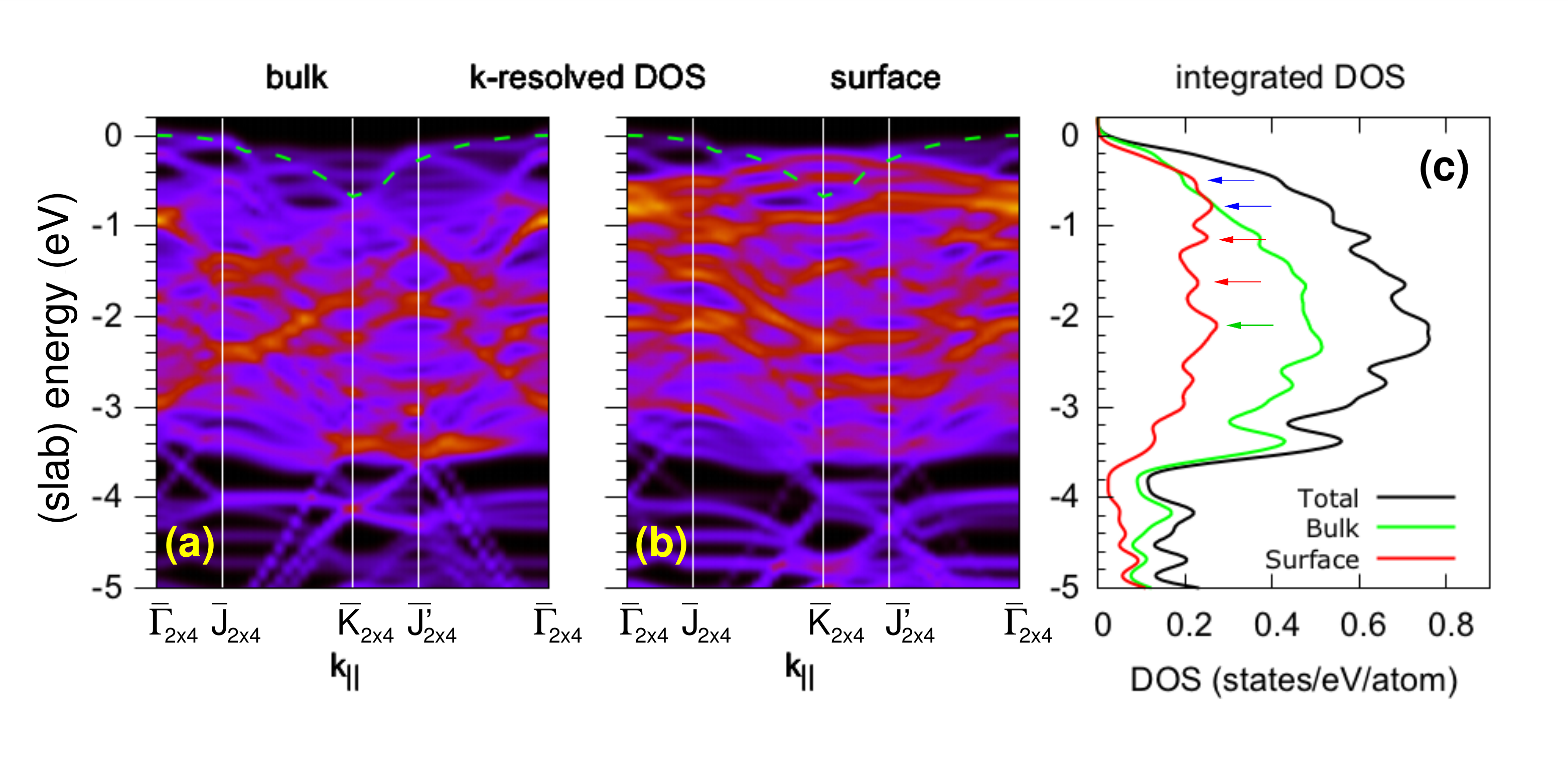}
\caption{Density of states of the GaAs(001) slab: k-resolved DOS around the $(2\times4)$ surface BZ for (a) bulk and (b) surface atoms only; (c) integrated DOS projected over surface, bulk, and whole slab. The upper limit of the projected bulk band structure is indicated by dashed lines in (a) and (b). %Panel (c) also shows the integrated DOS for primitive bulk GaAs, scaled such that the 3.4eV peaks align. 
}
\label{fig:GaAsBi-slab}
\end{figure*}
%%%%%%%%%%%%%%%%%%%%%%%%%%%%%

Finally, we consider the surface contribution to the electronic properties. Fig.~\ref{fig:GaAsBi-slab} shows the k-resolved DOS and integrated DOS of the $\beta2(2\times4)$ reconstructed slab, each projected onto surface and bulk regions as defined in Section.~\ref{sec:theory}. A number of `true' surface states, i.e., lying within the projected bulk gap, are visible around the $\Kbar_{2\times4}$ point as previously reported.\cite{Schmidt1996,Hogan2003b} 
These bands give rise to a peak in the total DOS at -0.5eV, as indicated by the blue arrow in (c). A second peak arising from surface-localized resonance states is observed at -0.8 eV, and is associated with a very flat band along $\Gbar$--$\Jbar_{2x4}$ in particular and an intense signal around $\Gbar$, as seen in panel (b). It likely corresponds to the flat feature (at -0.8eV) unfolded to $\Jbar$ in Fig.~\ref{fig:kPEEM-GaAs1}(h). 
We tentatively associate these two peaks with the SS shoulder identified in our UPS (Fig.~\ref{fig:UPS-GaAsBi}(a)) and the $S_1$ and $S_2$ surface states previously identified in ARPES data by Larsen et al~\cite{Larsen1982}.
Further peaks in the integrated surface DOS are observed at -1.2, -1.6, and -2.1eV. They arise from various resonances dispersing weakly along $\Gbar$--$\Jbar_{2x4}$ (compare with the unfolded bands in Fig.~\ref{fig:unfolded-GaAsBi}).
%and $\Jbar_{2x4}$--$\Kbar_{2x4}$, 
The latter states are thus well consistent with the $a_1$ and $a_2$ UPS features. 
We are not, however, able to make a clear interpretation of the $b$ feature, due to the considerable overlap with bulk states within the relevant energy range (2.5--3.5eV).

%%%%%%%%%%%%%%%%%%%%%%%%%%%%%%%%%%%%%%%%%%%%%%%%%%%%%
\section{Summary}
%%%%%%%%%%%%%%%%%%%%%%%%%%%%%%%%%%%%%%%%%%%%%%%%%%%%%%

We have studied structural and electronic properties of molecular beam epitaxy grown GaAs$_{1-x}$Bi$_{x}$ with varying Bi concentrations epilayers with (001) crystalline order and varying Bi content. Band structure properties of the well-known $\beta_2(2\times 4)$ reconstruction of pure GaAs(001) were measured by angle-resolved photoemission as well as photon energy dependent ultra-violet photoemission. By comparing the experimental data to density functional theory calculations of bulk GaAs and the $\beta_2(2\times4)$-reconstructed GaAs surface, we show that angle-resolved photoemission intensities at $21.2$~eV are governed by projections of bulk heavy and light hole bands in an extended Brillouin zone scheme along the high-symmetry directions X-W-X', X-U-$\Gamma$', $\Gamma$-K-X', and $\Gamma$-L-$\Gamma$, i.e. taking into account the $k_z$ dependence. From the {\it ab initio} slab calculation we attribute an experimentally observed (and previously unreported) dispersing band at binding energies of about 1~eV with a surface state. 

Bi atoms are found to be integrated as a Bi-rich surface layer and bulk Bi species which lead to a change of the surface order towards a $(1 \times 3)$ symmetry. We achieved Bi bulk concentrations of $\approx 1$ \% while preserving surface order and crystallinity high. In photoemission Bi-induced energy shifts and broadening effects of GaAs heavy and light hole bulk bands become evident, which are reproduced in respective calculations. Bi-induced electronic band structure effects underlines the potential of GaAs$_{1-x}$Bi$_{x}$  compounds for modulations in the optical band gap and thus optoelectronic applications.

%\begin{figure}
%\includegraphics[width=4.82278in,height=6.35820in]{media/image6.emf}
%\caption{bla}
%\end{figure}

%\includegraphics[width=6.30000in,height=3.80880in]{media/image7.emf}

\acknowledgments

This study was supported by the Bilateral Mobility Program between CNR (SAC.AD002.018.017) (Italy) and the Czech Academy of Sciences (CNR-16-02) and by the project Consolidate the Foundation 2015 - \textit{BILLY} - of the University of Rome Tor Vergata. C.H acknowledges CINECA under the ISCRA scheme for high-performance supercomputing resources and support.

\section*{Author Contributions}
Samples were grown and characterized by E.P. and F.A.; Photoemission and \emph{k}-PEEM measurements were performed by M.V., Y.P. and J.H. with the contribution of E.P., J.K. and F.A; C.H. performed all the theoretical calculations. The paper was written by J.H., C.H., E.P., and F.A., with the help and through contributions from all co-authors. All authors have given approval to the final version of the manuscript. The projects was initiated and conceptualized by J.H., E.P. and F.A.

%\bibliography{}

%\bibliography{Papers-GaAsBi} % Old working version
\bibliography{Projects-Active-Papers-GaAsBi}

\begin{thebibliography}{27}%
\makeatletter
\providecommand \@ifxundefined [1]{%
 \@ifx{#1\undefined}
}%
\providecommand \@ifnum [1]{%
 \ifnum #1\expandafter \@firstoftwo
 \else \expandafter \@secondoftwo
 \fi
}%
\providecommand \@ifx [1]{%
 \ifx #1\expandafter \@firstoftwo
 \else \expandafter \@secondoftwo
 \fi
}%
\providecommand \natexlab [1]{#1}%
\providecommand \enquote  [1]{``#1''}%
\providecommand \bibnamefont  [1]{#1}%
\providecommand \bibfnamefont [1]{#1}%
\providecommand \citenamefont [1]{#1}%
\providecommand \href@noop [0]{\@secondoftwo}%
\providecommand \href [0]{\begingroup \@sanitize@url \@href}%
\providecommand \@href[1]{\@@startlink{#1}\@@href}%
\providecommand \@@href[1]{\endgroup#1\@@endlink}%
\providecommand \@sanitize@url [0]{\catcode `\\12\catcode `\$12\catcode
  `\&12\catcode `\#12\catcode `\^12\catcode `\_12\catcode `\%12\relax}%
\providecommand \@@startlink[1]{}%
\providecommand \@@endlink[0]{}%
\providecommand \url  [0]{\begingroup\@sanitize@url \@url }%
\providecommand \@url [1]{\endgroup\@href {#1}{\urlprefix }}%
\providecommand \urlprefix  [0]{URL }%
\providecommand \Eprint [0]{\href }%
\providecommand \doibase [0]{http://dx.doi.org/}%
\providecommand \selectlanguage [0]{\@gobble}%
\providecommand \bibinfo  [0]{\@secondoftwo}%
\providecommand \bibfield  [0]{\@secondoftwo}%
\providecommand \translation [1]{[#1]}%
\providecommand \BibitemOpen [0]{}%
\providecommand \bibitemStop [0]{}%
\providecommand \bibitemNoStop [0]{.\EOS\space}%
\providecommand \EOS [0]{\spacefactor3000\relax}%
\providecommand \BibitemShut  [1]{\csname bibitem#1\endcsname}%
\let\auto@bib@innerbib\@empty
%</preamble>
\bibitem [{\citenamefont {Marko}\ \emph {et~al.}(2016)\citenamefont {Marko},
  \citenamefont {Broderick}, \citenamefont {Jin}, \citenamefont {Ludewig},
  \citenamefont {Stolz}, \citenamefont {Volz}, \citenamefont {Rorison},
  \citenamefont {O'Reilly},\ and\ \citenamefont {Sweeney}}]{Marko2016}%
  \BibitemOpen
  \bibfield  {author} {\bibinfo {author} {\bibfnamefont {Igor~P.}\ \bibnamefont
  {Marko}}, \bibinfo {author} {\bibfnamefont {Christopher~A.}\ \bibnamefont
  {Broderick}}, \bibinfo {author} {\bibfnamefont {Shirong}\ \bibnamefont
  {Jin}}, \bibinfo {author} {\bibfnamefont {Peter}\ \bibnamefont {Ludewig}},
  \bibinfo {author} {\bibfnamefont {Wolfgang}\ \bibnamefont {Stolz}}, \bibinfo
  {author} {\bibfnamefont {Kerstin}\ \bibnamefont {Volz}}, \bibinfo {author}
  {\bibfnamefont {Judy~M.}\ \bibnamefont {Rorison}}, \bibinfo {author}
  {\bibfnamefont {Eoin~P.}\ \bibnamefont {O'Reilly}}, \ and\ \bibinfo {author}
  {\bibfnamefont {Stephen~J.}\ \bibnamefont {Sweeney}},\ }\bibfield  {title}
  {\enquote {\bibinfo {title} {{Optical gain in GaAsBi/GaAs quantum well diode
  lasers}},}\ }\href {\doibase 10.1038/srep28863} {\bibfield  {journal}
  {\bibinfo  {journal} {Sci. Rep.}\ }\textbf {\bibinfo {volume} {6}},\ \bibinfo
  {pages} {28863} (\bibinfo {year} {2016})}\BibitemShut {NoStop}%
\bibitem [{\citenamefont {Richards}\ \emph {et~al.}(2015)\citenamefont
  {Richards}, \citenamefont {Bastiman}, \citenamefont {Roberts}, \citenamefont
  {Beanland}, \citenamefont {Walker},\ and\ \citenamefont
  {David}}]{Richards2015}%
  \BibitemOpen
  \bibfield  {author} {\bibinfo {author} {\bibfnamefont {Robert~D.}\
  \bibnamefont {Richards}}, \bibinfo {author} {\bibfnamefont {Faebian}\
  \bibnamefont {Bastiman}}, \bibinfo {author} {\bibfnamefont {John~S.}\
  \bibnamefont {Roberts}}, \bibinfo {author} {\bibfnamefont {Richard}\
  \bibnamefont {Beanland}}, \bibinfo {author} {\bibfnamefont {David}\
  \bibnamefont {Walker}}, \ and\ \bibinfo {author} {\bibfnamefont {John P~R}\
  \bibnamefont {David}},\ }\bibfield  {title} {\enquote {\bibinfo {title} {{MBE
  grown GaAsBi/GaAs multiple quantum well structures: Structural and optical
  characterization}},}\ }\href {\doibase 10.1016/j.jcrysgro.2015.02.053}
  {\bibfield  {journal} {\bibinfo  {journal} {J. Cryst. Growth}\ }\textbf
  {\bibinfo {volume} {425}},\ \bibinfo {pages} {237--240} (\bibinfo {year}
  {2015})}\BibitemShut {NoStop}%
\bibitem [{\citenamefont {Yamashita}\ \emph {et~al.}(2006)\citenamefont
  {Yamashita}, \citenamefont {Yoshimoto},\ and\ \citenamefont
  {Oe}}]{Yamashita2006}%
  \BibitemOpen
  \bibfield  {author} {\bibinfo {author} {\bibfnamefont {K.}~\bibnamefont
  {Yamashita}}, \bibinfo {author} {\bibfnamefont {M.}~\bibnamefont
  {Yoshimoto}}, \ and\ \bibinfo {author} {\bibfnamefont {K.}~\bibnamefont
  {Oe}},\ }\bibfield  {title} {\enquote {\bibinfo {title}
  {{Temperature-insensitive refractive index of GaAsBi alloy for laser diode in
  WDM optical communication}},}\ }\href {\doibase 10.1002/pssc.200564110}
  {\bibfield  {journal} {\bibinfo  {journal} {Phys. status solidi}\ }\textbf
  {\bibinfo {volume} {3}},\ \bibinfo {pages} {693--696} (\bibinfo {year}
  {2006})}\BibitemShut {NoStop}%
\bibitem [{\citenamefont {Francoeur}\ \emph {et~al.}(2003)\citenamefont
  {Francoeur}, \citenamefont {Seong}, \citenamefont {Mascarenhas},
  \citenamefont {Tixier}, \citenamefont {Adamcyk},\ and\ \citenamefont
  {Tiedje}}]{Francoeur2003}%
  \BibitemOpen
  \bibfield  {author} {\bibinfo {author} {\bibfnamefont {S.}~\bibnamefont
  {Francoeur}}, \bibinfo {author} {\bibfnamefont {M.-J.}\ \bibnamefont
  {Seong}}, \bibinfo {author} {\bibfnamefont {A.}~\bibnamefont {Mascarenhas}},
  \bibinfo {author} {\bibfnamefont {S.}~\bibnamefont {Tixier}}, \bibinfo
  {author} {\bibfnamefont {M.}~\bibnamefont {Adamcyk}}, \ and\ \bibinfo
  {author} {\bibfnamefont {T.}~\bibnamefont {Tiedje}},\ }\bibfield  {title}
  {\enquote {\bibinfo {title} {{Band gap of GaAs1−xBix,
  0{\textless}x{\textless}3.6{\%}}},}\ }\href {\doibase 10.1063/1.1581983}
  {\bibfield  {journal} {\bibinfo  {journal} {Appl. Phys. Lett.}\ }\textbf
  {\bibinfo {volume} {82}},\ \bibinfo {pages} {3874--3876} (\bibinfo {year}
  {2003})}\BibitemShut {NoStop}%
\bibitem [{\citenamefont {Kini}\ \emph {et~al.}(2009)\citenamefont {Kini},
  \citenamefont {Bhusal}, \citenamefont {Ptak}, \citenamefont {France},\ and\
  \citenamefont {Mascarenhas}}]{Kini2009}%
  \BibitemOpen
  \bibfield  {author} {\bibinfo {author} {\bibfnamefont {R.~N.}\ \bibnamefont
  {Kini}}, \bibinfo {author} {\bibfnamefont {L.}~\bibnamefont {Bhusal}},
  \bibinfo {author} {\bibfnamefont {A.~J.}\ \bibnamefont {Ptak}}, \bibinfo
  {author} {\bibfnamefont {R.}~\bibnamefont {France}}, \ and\ \bibinfo {author}
  {\bibfnamefont {A.}~\bibnamefont {Mascarenhas}},\ }\bibfield  {title}
  {\enquote {\bibinfo {title} {{Electron Hall mobility in GaAsBi}},}\ }\href
  {\doibase 10.1063/1.3204670} {\bibfield  {journal} {\bibinfo  {journal} {J.
  Appl. Phys.}\ }\textbf {\bibinfo {volume} {106}},\ \bibinfo {pages} {043705}
  (\bibinfo {year} {2009})}\BibitemShut {NoStop}%
\bibitem [{\citenamefont {Bastiman}\ \emph {et~al.}(2012)\citenamefont
  {Bastiman}, \citenamefont {Cullis}, \citenamefont {David},\ and\
  \citenamefont {Sweeney}}]{Bastiman2012}%
  \BibitemOpen
  \bibfield  {author} {\bibinfo {author} {\bibfnamefont {F.}~\bibnamefont
  {Bastiman}}, \bibinfo {author} {\bibfnamefont {A.~G.}\ \bibnamefont
  {Cullis}}, \bibinfo {author} {\bibfnamefont {J.~P~R}\ \bibnamefont {David}},
  \ and\ \bibinfo {author} {\bibfnamefont {S.~J.}\ \bibnamefont {Sweeney}},\
  }\bibfield  {title} {\enquote {\bibinfo {title} {{Bi incorporation in
  GaAs(100)-2??1 and 4??3 reconstructions investigated by RHEED and STM}},}\
  }\href {\doibase 10.1016/j.jcrysgro.2011.12.058} {\bibfield  {journal}
  {\bibinfo  {journal} {J. Cryst. Growth}\ }\textbf {\bibinfo {volume} {341}},\
  \bibinfo {pages} {19--23} (\bibinfo {year} {2012})}\BibitemShut {NoStop}%
\bibitem [{\citenamefont {Kr{\"{o}}mker}\ \emph {et~al.}(2008)\citenamefont
  {Kr{\"{o}}mker}, \citenamefont {Escher}, \citenamefont {Funnemann},
  \citenamefont {Hartung}, \citenamefont {Engelhard},\ and\ \citenamefont
  {Kirschner}}]{Kromker2008}%
  \BibitemOpen
  \bibfield  {author} {\bibinfo {author} {\bibfnamefont {B.}~\bibnamefont
  {Kr{\"{o}}mker}}, \bibinfo {author} {\bibfnamefont {M.}~\bibnamefont
  {Escher}}, \bibinfo {author} {\bibfnamefont {D.}~\bibnamefont {Funnemann}},
  \bibinfo {author} {\bibfnamefont {D.}~\bibnamefont {Hartung}}, \bibinfo
  {author} {\bibfnamefont {H.}~\bibnamefont {Engelhard}}, \ and\ \bibinfo
  {author} {\bibfnamefont {J.}~\bibnamefont {Kirschner}},\ }\bibfield  {title}
  {\enquote {\bibinfo {title} {{Development of a momentum microscope for time
  resolved band structure imaging}},}\ }\href {\doibase 10.1063/1.2918133}
  {\bibfield  {journal} {\bibinfo  {journal} {Rev. Sci. Instrum.}\ }\textbf
  {\bibinfo {volume} {79}},\ \bibinfo {pages} {053702} (\bibinfo {year}
  {2008})}\BibitemShut {NoStop}%
\bibitem [{\citenamefont {Giannozzi}\ \emph {et~al.}(2017)\citenamefont
  {Giannozzi}, \citenamefont {Andreussi}, \citenamefont {Brumme}, \citenamefont
  {Bunau}, \citenamefont {{Buongiorno Nardelli}}, \citenamefont {Calandra},
  \citenamefont {Car}, \citenamefont {Cavazzoni}, \citenamefont {Ceresoli},
  \citenamefont {Cococcioni}, \citenamefont {Colonna}, \citenamefont
  {Carnimeo}, \citenamefont {{Dal Corso}}, \citenamefont {de~Gironcoli},
  \citenamefont {Delugas}, \citenamefont {DiStasio}, \citenamefont {Ferretti},
  \citenamefont {Floris}, \citenamefont {Fratesi}, \citenamefont {Fugallo},
  \citenamefont {Gebauer}, \citenamefont {Gerstmann}, \citenamefont {Giustino},
  \citenamefont {Gorni}, \citenamefont {Jia}, \citenamefont {Kawamura},
  \citenamefont {Ko}, \citenamefont {Kokalj}, \citenamefont
  {K{\"{u}}{\c{c}}{\"{u}}kbenli}, \citenamefont {Lazzeri}, \citenamefont
  {Marsili}, \citenamefont {Marzari}, \citenamefont {Mauri}, \citenamefont
  {Nguyen}, \citenamefont {Nguyen}, \citenamefont {Otero-de-la Roza},
  \citenamefont {Paulatto}, \citenamefont {Ponc{\'{e}}}, \citenamefont {Rocca},
  \citenamefont {Sabatini}, \citenamefont {Santra}, \citenamefont {Schlipf},
  \citenamefont {Seitsonen}, \citenamefont {Smogunov}, \citenamefont {Timrov},
  \citenamefont {Thonhauser}, \citenamefont {Umari}, \citenamefont {Vast},
  \citenamefont {Wu},\ and\ \citenamefont {Baroni}}]{Giannozzi2017}%
  \BibitemOpen
  \bibfield  {author} {\bibinfo {author} {\bibfnamefont {P}~\bibnamefont
  {Giannozzi}}, \bibinfo {author} {\bibfnamefont {O}~\bibnamefont {Andreussi}},
  \bibinfo {author} {\bibfnamefont {T}~\bibnamefont {Brumme}}, \bibinfo
  {author} {\bibfnamefont {O}~\bibnamefont {Bunau}}, \bibinfo {author}
  {\bibfnamefont {M}~\bibnamefont {{Buongiorno Nardelli}}}, \bibinfo {author}
  {\bibfnamefont {M}~\bibnamefont {Calandra}}, \bibinfo {author} {\bibfnamefont
  {R}~\bibnamefont {Car}}, \bibinfo {author} {\bibfnamefont {C}~\bibnamefont
  {Cavazzoni}}, \bibinfo {author} {\bibfnamefont {D}~\bibnamefont {Ceresoli}},
  \bibinfo {author} {\bibfnamefont {M}~\bibnamefont {Cococcioni}}, \bibinfo
  {author} {\bibfnamefont {N}~\bibnamefont {Colonna}}, \bibinfo {author}
  {\bibfnamefont {I}~\bibnamefont {Carnimeo}}, \bibinfo {author} {\bibfnamefont
  {A}~\bibnamefont {{Dal Corso}}}, \bibinfo {author} {\bibfnamefont
  {S}~\bibnamefont {de~Gironcoli}}, \bibinfo {author} {\bibfnamefont
  {P}~\bibnamefont {Delugas}}, \bibinfo {author} {\bibfnamefont {R~A}\
  \bibnamefont {DiStasio}}, \bibinfo {author} {\bibfnamefont {A}~\bibnamefont
  {Ferretti}}, \bibinfo {author} {\bibfnamefont {A}~\bibnamefont {Floris}},
  \bibinfo {author} {\bibfnamefont {G}~\bibnamefont {Fratesi}}, \bibinfo
  {author} {\bibfnamefont {G}~\bibnamefont {Fugallo}}, \bibinfo {author}
  {\bibfnamefont {R}~\bibnamefont {Gebauer}}, \bibinfo {author} {\bibfnamefont
  {U}~\bibnamefont {Gerstmann}}, \bibinfo {author} {\bibfnamefont
  {F}~\bibnamefont {Giustino}}, \bibinfo {author} {\bibfnamefont
  {T}~\bibnamefont {Gorni}}, \bibinfo {author} {\bibfnamefont {J}~\bibnamefont
  {Jia}}, \bibinfo {author} {\bibfnamefont {M}~\bibnamefont {Kawamura}},
  \bibinfo {author} {\bibfnamefont {H-Y}\ \bibnamefont {Ko}}, \bibinfo {author}
  {\bibfnamefont {A}~\bibnamefont {Kokalj}}, \bibinfo {author} {\bibfnamefont
  {E}~\bibnamefont {K{\"{u}}{\c{c}}{\"{u}}kbenli}}, \bibinfo {author}
  {\bibfnamefont {M}~\bibnamefont {Lazzeri}}, \bibinfo {author} {\bibfnamefont
  {M}~\bibnamefont {Marsili}}, \bibinfo {author} {\bibfnamefont
  {N}~\bibnamefont {Marzari}}, \bibinfo {author} {\bibfnamefont
  {F}~\bibnamefont {Mauri}}, \bibinfo {author} {\bibfnamefont {N~L}\
  \bibnamefont {Nguyen}}, \bibinfo {author} {\bibfnamefont {H-V}\ \bibnamefont
  {Nguyen}}, \bibinfo {author} {\bibfnamefont {A}~\bibnamefont {Otero-de-la
  Roza}}, \bibinfo {author} {\bibfnamefont {L}~\bibnamefont {Paulatto}},
  \bibinfo {author} {\bibfnamefont {S}~\bibnamefont {Ponc{\'{e}}}}, \bibinfo
  {author} {\bibfnamefont {D}~\bibnamefont {Rocca}}, \bibinfo {author}
  {\bibfnamefont {R}~\bibnamefont {Sabatini}}, \bibinfo {author} {\bibfnamefont
  {B}~\bibnamefont {Santra}}, \bibinfo {author} {\bibfnamefont {M}~\bibnamefont
  {Schlipf}}, \bibinfo {author} {\bibfnamefont {A~P}\ \bibnamefont
  {Seitsonen}}, \bibinfo {author} {\bibfnamefont {A}~\bibnamefont {Smogunov}},
  \bibinfo {author} {\bibfnamefont {I}~\bibnamefont {Timrov}}, \bibinfo
  {author} {\bibfnamefont {T}~\bibnamefont {Thonhauser}}, \bibinfo {author}
  {\bibfnamefont {P}~\bibnamefont {Umari}}, \bibinfo {author} {\bibfnamefont
  {N}~\bibnamefont {Vast}}, \bibinfo {author} {\bibfnamefont {X}~\bibnamefont
  {Wu}}, \ and\ \bibinfo {author} {\bibfnamefont {S}~\bibnamefont {Baroni}},\
  }\bibfield  {title} {\enquote {\bibinfo {title} {{Advanced capabilities for
  materials modelling with Quantum ESPRESSO}},}\ }\href {\doibase
  10.1088/1361-648X/aa8f79} {\bibfield  {journal} {\bibinfo  {journal} {J.
  Phys. Condens. Matter}\ }\textbf {\bibinfo {volume} {29}},\ \bibinfo {pages}
  {465901} (\bibinfo {year} {2017})}\BibitemShut {NoStop}%
\bibitem [{\citenamefont {{Dal Corso}}\ and\ \citenamefont
  {Conte}(2005)}]{DalCorso2005}%
  \BibitemOpen
  \bibfield  {author} {\bibinfo {author} {\bibfnamefont {Andrea}\ \bibnamefont
  {{Dal Corso}}}\ and\ \bibinfo {author} {\bibfnamefont {Adriano~Mosca}\
  \bibnamefont {Conte}},\ }\bibfield  {title} {\enquote {\bibinfo {title}
  {{Spin-orbit coupling with ultrasoft pseudopotentials: Application to Au and
  Pt}},}\ }\href {\doibase 10.1103/PhysRevB.71.115106} {\bibfield  {journal}
  {\bibinfo  {journal} {Phys. Rev. B - Condens. Matter Mater. Phys.}\ }\textbf
  {\bibinfo {volume} {71}},\ \bibinfo {pages} {115106} (\bibinfo {year}
  {2005})}\BibitemShut {NoStop}%
\bibitem [{\citenamefont {Rappe}\ \emph {et~al.}(1990)\citenamefont {Rappe},
  \citenamefont {Rabe}, \citenamefont {Kaxiras},\ and\ \citenamefont
  {Joannopoulos}}]{Rappe1990}%
  \BibitemOpen
  \bibfield  {author} {\bibinfo {author} {\bibfnamefont {Andrew~M.}\
  \bibnamefont {Rappe}}, \bibinfo {author} {\bibfnamefont {Karin~M.}\
  \bibnamefont {Rabe}}, \bibinfo {author} {\bibfnamefont {Efthimios}\
  \bibnamefont {Kaxiras}}, \ and\ \bibinfo {author} {\bibfnamefont {J.~D.}\
  \bibnamefont {Joannopoulos}},\ }\bibfield  {title} {\enquote {\bibinfo
  {title} {{Optimized pseudopotentials}},}\ }\href {\doibase
  10.1103/PhysRevB.41.1227} {\bibfield  {journal} {\bibinfo  {journal} {Phys.
  Rev. B}\ }\textbf {\bibinfo {volume} {41}},\ \bibinfo {pages} {1227--1230}
  (\bibinfo {year} {1990})}\BibitemShut {NoStop}%
\bibitem [{\citenamefont {Medeiros}\ \emph {et~al.}(2014)\citenamefont
  {Medeiros}, \citenamefont {Stafstr{\"{o}}m},\ and\ \citenamefont
  {Bj{\"{o}}rk}}]{Medeiros2014}%
  \BibitemOpen
  \bibfield  {author} {\bibinfo {author} {\bibfnamefont {Paulo V~C}\
  \bibnamefont {Medeiros}}, \bibinfo {author} {\bibfnamefont {Sven}\
  \bibnamefont {Stafstr{\"{o}}m}}, \ and\ \bibinfo {author} {\bibfnamefont
  {Jonas}\ \bibnamefont {Bj{\"{o}}rk}},\ }\bibfield  {title} {\enquote
  {\bibinfo {title} {{Effects of extrinsic and intrinsic perturbations on the
  electronic structure of graphene: Retaining an effective primitive cell band
  structure by band unfolding}},}\ }\href {\doibase 10.1103/PhysRevB.89.041407}
  {\bibfield  {journal} {\bibinfo  {journal} {Phys. Rev. B - Condens. Matter
  Mater. Phys.}\ }\textbf {\bibinfo {volume} {89}} (\bibinfo {year} {2014}),\
  10.1103/PhysRevB.89.041407}\BibitemShut {NoStop}%
\bibitem [{\citenamefont {Medeiros}\ \emph {et~al.}(2015)\citenamefont
  {Medeiros}, \citenamefont {Tsirkin}, \citenamefont {Stafstr??m},\ and\
  \citenamefont {Bj??rk}}]{Medeiros2015}%
  \BibitemOpen
  \bibfield  {author} {\bibinfo {author} {\bibfnamefont {Paulo V~C}\
  \bibnamefont {Medeiros}}, \bibinfo {author} {\bibfnamefont {Stepan~S.}\
  \bibnamefont {Tsirkin}}, \bibinfo {author} {\bibfnamefont {Sven}\
  \bibnamefont {Stafstr??m}}, \ and\ \bibinfo {author} {\bibfnamefont {Jonas}\
  \bibnamefont {Bj??rk}},\ }\bibfield  {title} {\enquote {\bibinfo {title}
  {{Unfolding spinor wave functions and expectation values of general
  operators: Introducing the unfolding-density operator}},}\ }\href {\doibase
  10.1103/PhysRevB.91.041116} {\bibfield  {journal} {\bibinfo  {journal} {Phys.
  Rev. B - Condens. Matter Mater. Phys.}\ }\textbf {\bibinfo {volume} {91}},\
  \bibinfo {pages} {1--5} (\bibinfo {year} {2015})},\ \Eprint
  {http://arxiv.org/abs/1409.5343} {arXiv:1409.5343} \BibitemShut {NoStop}%
\bibitem [{\citenamefont {Kraut}\ \emph {et~al.}(1980)\citenamefont {Kraut},
  \citenamefont {Grant}, \citenamefont {Waldrop},\ and\ \citenamefont
  {Kowalczyk}}]{Kraut1980}%
  \BibitemOpen
  \bibfield  {author} {\bibinfo {author} {\bibfnamefont {E.~A.}\ \bibnamefont
  {Kraut}}, \bibinfo {author} {\bibfnamefont {R.~W.}\ \bibnamefont {Grant}},
  \bibinfo {author} {\bibfnamefont {J.~R.}\ \bibnamefont {Waldrop}}, \ and\
  \bibinfo {author} {\bibfnamefont {S.~P.}\ \bibnamefont {Kowalczyk}},\
  }\bibfield  {title} {\enquote {\bibinfo {title} {{Precise Determination of
  the Valence-Band Edge in X-Ray Photoemission Spectra: Application to
  Measurement of Semiconductor Interface Potentials}},}\ }\href {\doibase
  10.1103/PhysRevLett.44.1620} {\bibfield  {journal} {\bibinfo  {journal}
  {Phys. Rev. Lett.}\ }\textbf {\bibinfo {volume} {44}},\ \bibinfo {pages}
  {1620--1623} (\bibinfo {year} {1980})}\BibitemShut {NoStop}%
\bibitem [{\citenamefont {Laukkanen}\ \emph {et~al.}(2017)\citenamefont
  {Laukkanen}, \citenamefont {Punkkinen}, \citenamefont {Lahti}, \citenamefont
  {Puustinen}, \citenamefont {Tuominen}, \citenamefont {Hilska}, \citenamefont
  {M{\"{a}}kel{\"{a}}}, \citenamefont {Dahl}, \citenamefont {Yasir},
  \citenamefont {Kuzmin}, \citenamefont {Osiecki}, \citenamefont {Schulte},
  \citenamefont {Guina},\ and\ \citenamefont {Kokko}}]{Laukkanen2017}%
  \BibitemOpen
  \bibfield  {author} {\bibinfo {author} {\bibfnamefont {P.}~\bibnamefont
  {Laukkanen}}, \bibinfo {author} {\bibfnamefont {M.P.J.}\ \bibnamefont
  {Punkkinen}}, \bibinfo {author} {\bibfnamefont {A.}~\bibnamefont {Lahti}},
  \bibinfo {author} {\bibfnamefont {J.}~\bibnamefont {Puustinen}}, \bibinfo
  {author} {\bibfnamefont {M.}~\bibnamefont {Tuominen}}, \bibinfo {author}
  {\bibfnamefont {J.}~\bibnamefont {Hilska}}, \bibinfo {author} {\bibfnamefont
  {J.}~\bibnamefont {M{\"{a}}kel{\"{a}}}}, \bibinfo {author} {\bibfnamefont
  {J.}~\bibnamefont {Dahl}}, \bibinfo {author} {\bibfnamefont {M.}~\bibnamefont
  {Yasir}}, \bibinfo {author} {\bibfnamefont {M.}~\bibnamefont {Kuzmin}},
  \bibinfo {author} {\bibfnamefont {J.R.}\ \bibnamefont {Osiecki}}, \bibinfo
  {author} {\bibfnamefont {K.}~\bibnamefont {Schulte}}, \bibinfo {author}
  {\bibfnamefont {M.}~\bibnamefont {Guina}}, \ and\ \bibinfo {author}
  {\bibfnamefont {K.}~\bibnamefont {Kokko}},\ }\bibfield  {title} {\enquote
  {\bibinfo {title} {{Local variation in Bi crystal sites of epitaxial GaAsBi
  studied by photoelectron spectroscopy and first-principles calculations}},}\
  }\href {\doibase 10.1016/j.apsusc.2016.11.009} {\bibfield  {journal}
  {\bibinfo  {journal} {Appl. Surf. Sci.}\ }\textbf {\bibinfo {volume} {396}},\
  \bibinfo {pages} {688--694} (\bibinfo {year} {2017})}\BibitemShut {NoStop}%
\bibitem [{\citenamefont {Sales}\ \emph {et~al.}(2011)\citenamefont {Sales},
  \citenamefont {Guerrero}, \citenamefont {Rodrigo}, \citenamefont {Galindo},
  \citenamefont {Y{\'{a}}{\~{n}}ez}, \citenamefont {Shafi}, \citenamefont
  {Khatab}, \citenamefont {Mari}, \citenamefont {Henini}, \citenamefont
  {Novikov}, \citenamefont {Chisholm},\ and\ \citenamefont
  {Molina}}]{Sales2011}%
  \BibitemOpen
  \bibfield  {author} {\bibinfo {author} {\bibfnamefont {D.~L.}\ \bibnamefont
  {Sales}}, \bibinfo {author} {\bibfnamefont {E.}~\bibnamefont {Guerrero}},
  \bibinfo {author} {\bibfnamefont {J.~F.}\ \bibnamefont {Rodrigo}}, \bibinfo
  {author} {\bibfnamefont {P.~L.}\ \bibnamefont {Galindo}}, \bibinfo {author}
  {\bibfnamefont {A.}~\bibnamefont {Y{\'{a}}{\~{n}}ez}}, \bibinfo {author}
  {\bibfnamefont {M.}~\bibnamefont {Shafi}}, \bibinfo {author} {\bibfnamefont
  {A.}~\bibnamefont {Khatab}}, \bibinfo {author} {\bibfnamefont {R.~H.}\
  \bibnamefont {Mari}}, \bibinfo {author} {\bibfnamefont {M.}~\bibnamefont
  {Henini}}, \bibinfo {author} {\bibfnamefont {S.}~\bibnamefont {Novikov}},
  \bibinfo {author} {\bibfnamefont {M.~F.}\ \bibnamefont {Chisholm}}, \ and\
  \bibinfo {author} {\bibfnamefont {S.~I.}\ \bibnamefont {Molina}},\ }\bibfield
   {title} {\enquote {\bibinfo {title} {{Distribution of bismuth atoms in
  epitaxial GaAsBi}},}\ }\href {\doibase 10.1063/1.3562376} {\bibfield
  {journal} {\bibinfo  {journal} {Appl. Phys. Lett.}\ }\textbf {\bibinfo
  {volume} {98}},\ \bibinfo {pages} {101902} (\bibinfo {year}
  {2011})}\BibitemShut {NoStop}%
\bibitem [{\citenamefont {Cai}\ \emph {et~al.}(1992)\citenamefont {Cai},
  \citenamefont {Stampfl}, \citenamefont {Riley}, \citenamefont {Leckey},
  \citenamefont {Usher},\ and\ \citenamefont {Ley}}]{Cai1992}%
  \BibitemOpen
  \bibfield  {author} {\bibinfo {author} {\bibfnamefont {Y.~Q.}\ \bibnamefont
  {Cai}}, \bibinfo {author} {\bibfnamefont {A.~P.~J.}\ \bibnamefont {Stampfl}},
  \bibinfo {author} {\bibfnamefont {J.~D.}\ \bibnamefont {Riley}}, \bibinfo
  {author} {\bibfnamefont {R.~C.~G.}\ \bibnamefont {Leckey}}, \bibinfo {author}
  {\bibfnamefont {B.}~\bibnamefont {Usher}}, \ and\ \bibinfo {author}
  {\bibfnamefont {L.}~\bibnamefont {Ley}},\ }\bibfield  {title} {\enquote
  {\bibinfo {title} {{Two-dimensional electronic structure {\textless}math
  display="inline"{\textgreater} {\textless}mrow{\textgreater}
  {\textless}msub{\textgreater} {\textless}mrow{\textgreater} {\textless}mi
  mathvariant="italic"{\textgreater}E{\textless}/mi{\textgreater}
  {\textless}/mrow{\textgreater} {\textless}mrow{\textgreater} {\textless}mi
  mathvariant="italic"{\textgreater}i{\textless}/mi{\textgreater}
  {\textless}/mrow{\textgreater} {\textless}/msub{\textgreater}
  {\textless}/mrow{\textgreater} {\textless}/math{\textgreater} (
  {\textless}math display="inline"{\textgreater} {\textless}mrow{\textgreater}
  {\textless}msubsup{\textgreater} {\textless}mrow{\textgreater} {\textless}mi
  mathvari}},}\ }\href {\doibase 10.1103/PhysRevB.46.6891} {\bibfield
  {journal} {\bibinfo  {journal} {Phys. Rev. B}\ }\textbf {\bibinfo {volume}
  {46}},\ \bibinfo {pages} {6891--6901} (\bibinfo {year} {1992})}\BibitemShut
  {NoStop}%
\bibitem [{\citenamefont {Chiang}\ \emph {et~al.}(1983)\citenamefont {Chiang},
  \citenamefont {Ludeke},\ and\ \citenamefont {Aono}}]{Chiang1983}%
  \BibitemOpen
  \bibfield  {author} {\bibinfo {author} {\bibfnamefont {TC}~\bibnamefont
  {Chiang}}, \bibinfo {author} {\bibfnamefont {R}~\bibnamefont {Ludeke}}, \
  and\ \bibinfo {author} {\bibfnamefont {M}~\bibnamefont {Aono}},\ }\bibfield
  {title} {\enquote {\bibinfo {title} {{Angle-resolved photoemission studies of
  GaAs (100) surfaces grown by molecular-beam epitaxy}},}\ }\href
  {http://prb.aps.org/abstract/PRB/v27/i8/p4770{\_}1} {\bibfield  {journal}
  {\bibinfo  {journal} {Phys. Rev. B}\ }\textbf {\bibinfo {volume} {27}},\
  \bibinfo {pages} {4770--4778} (\bibinfo {year} {1983})}\BibitemShut {NoStop}%
\bibitem [{\citenamefont {Olde}\ \emph {et~al.}(1990)\citenamefont {Olde},
  \citenamefont {Mante}, \citenamefont {Barnscheidt}, \citenamefont {Kipp},
  \citenamefont {Kuhr}, \citenamefont {Manzke}, \citenamefont {Skibowski},
  \citenamefont {Henk},\ and\ \citenamefont {Schattke}}]{Olde1990}%
  \BibitemOpen
  \bibfield  {author} {\bibinfo {author} {\bibfnamefont {J.}~\bibnamefont
  {Olde}}, \bibinfo {author} {\bibfnamefont {G.}~\bibnamefont {Mante}},
  \bibinfo {author} {\bibfnamefont {H.-P.}\ \bibnamefont {Barnscheidt}},
  \bibinfo {author} {\bibfnamefont {L.}~\bibnamefont {Kipp}}, \bibinfo {author}
  {\bibfnamefont {J.-C.}\ \bibnamefont {Kuhr}}, \bibinfo {author}
  {\bibfnamefont {R.}~\bibnamefont {Manzke}}, \bibinfo {author} {\bibfnamefont
  {M.}~\bibnamefont {Skibowski}}, \bibinfo {author} {\bibfnamefont
  {J.}~\bibnamefont {Henk}}, \ and\ \bibinfo {author} {\bibfnamefont
  {W.}~\bibnamefont {Schattke}},\ }\bibfield  {title} {\enquote {\bibinfo
  {title} {{Electronic structure of GaAs(001)}},}\ }\href {\doibase
  10.1103/PhysRevB.41.9958} {\bibfield  {journal} {\bibinfo  {journal} {Phys.
  Rev. B}\ }\textbf {\bibinfo {volume} {41}},\ \bibinfo {pages} {9958--9965}
  (\bibinfo {year} {1990})},\ \Eprint {http://arxiv.org/abs/arXiv:1107.0075v1}
  {arXiv:arXiv:1107.0075v1} \BibitemShut {NoStop}%
\bibitem [{\citenamefont {Larsen}\ and\ \citenamefont {van~der
  Veen}(1982)}]{Larsen1982}%
  \BibitemOpen
  \bibfield  {author} {\bibinfo {author} {\bibfnamefont {P~K}\ \bibnamefont
  {Larsen}}\ and\ \bibinfo {author} {\bibfnamefont {J~D}\ \bibnamefont {van~der
  Veen}},\ }\bibfield  {title} {\enquote {\bibinfo {title} {{Surface band
  structure of MBE-grown GaAs(001)-2 × 4}},}\ }\href {\doibase
  10.1088/0022-3719/15/13/010} {\bibfield  {journal} {\bibinfo  {journal} {J.
  Phys. C Solid State Phys.}\ }\textbf {\bibinfo {volume} {15}},\ \bibinfo
  {pages} {L431--L435} (\bibinfo {year} {1982})}\BibitemShut {NoStop}%
\bibitem [{\citenamefont {Placidi}\ \emph {et~al.}(2006)\citenamefont
  {Placidi}, \citenamefont {Hogan}, \citenamefont {Arciprete}, \citenamefont
  {Fanfoni}, \citenamefont {Patella}, \citenamefont {{Del Sole}},\ and\
  \citenamefont {Balzarotti}}]{Placidi2006}%
  \BibitemOpen
  \bibfield  {author} {\bibinfo {author} {\bibfnamefont {E.}~\bibnamefont
  {Placidi}}, \bibinfo {author} {\bibfnamefont {C.}~\bibnamefont {Hogan}},
  \bibinfo {author} {\bibfnamefont {F.}~\bibnamefont {Arciprete}}, \bibinfo
  {author} {\bibfnamefont {M.}~\bibnamefont {Fanfoni}}, \bibinfo {author}
  {\bibfnamefont {F.}~\bibnamefont {Patella}}, \bibinfo {author} {\bibfnamefont
  {R.}~\bibnamefont {{Del Sole}}}, \ and\ \bibinfo {author} {\bibfnamefont
  {A.}~\bibnamefont {Balzarotti}},\ }\bibfield  {title} {\enquote {\bibinfo
  {title} {{Adsorption of molecular oxygen on GaAs(001) studied using
  high-resolution electron energy-loss spectroscopy}},}\ }\href {\doibase
  10.1103/PhysRevB.73.205345} {\bibfield  {journal} {\bibinfo  {journal} {Phys.
  Rev. B}\ }\textbf {\bibinfo {volume} {73}},\ \bibinfo {pages} {205345}
  (\bibinfo {year} {2006})}\BibitemShut {NoStop}%
\bibitem [{\citenamefont {Arciprete}\ \emph {et~al.}(2004)\citenamefont
  {Arciprete}, \citenamefont {Goletti}, \citenamefont {Placidi}, \citenamefont
  {Hogan}, \citenamefont {Chiaradia}, \citenamefont {Fanfoni}, \citenamefont
  {Patella},\ and\ \citenamefont {Balzarotti}}]{Arciprete2004a}%
  \BibitemOpen
  \bibfield  {author} {\bibinfo {author} {\bibfnamefont {F.}~\bibnamefont
  {Arciprete}}, \bibinfo {author} {\bibfnamefont {C.}~\bibnamefont {Goletti}},
  \bibinfo {author} {\bibfnamefont {E.}~\bibnamefont {Placidi}}, \bibinfo
  {author} {\bibfnamefont {C.}~\bibnamefont {Hogan}}, \bibinfo {author}
  {\bibfnamefont {P.}~\bibnamefont {Chiaradia}}, \bibinfo {author}
  {\bibfnamefont {M.}~\bibnamefont {Fanfoni}}, \bibinfo {author} {\bibfnamefont
  {F.}~\bibnamefont {Patella}}, \ and\ \bibinfo {author} {\bibfnamefont
  {A.}~\bibnamefont {Balzarotti}},\ }\bibfield  {title} {\enquote {\bibinfo
  {title} {{Surface states at the GaAs(001)2x4 surface}},}\ }\href {\doibase
  10.1103/PhysRevB.69.081308} {\bibfield  {journal} {\bibinfo  {journal} {Phys.
  Rev. B}\ }\textbf {\bibinfo {volume} {69}},\ \bibinfo {pages} {081308}
  (\bibinfo {year} {2004})}\BibitemShut {NoStop}%
\bibitem [{\citenamefont {Souma}\ \emph {et~al.}(2016)\citenamefont {Souma},
  \citenamefont {Chen}, \citenamefont {Oszwa{\l}dowski}, \citenamefont {Sato},
  \citenamefont {Matsukura}, \citenamefont {Dietl}, \citenamefont {Ohno},\ and\
  \citenamefont {Takahashi}}]{Souma2016}%
  \BibitemOpen
  \bibfield  {author} {\bibinfo {author} {\bibfnamefont {S.}~\bibnamefont
  {Souma}}, \bibinfo {author} {\bibfnamefont {L.}~\bibnamefont {Chen}},
  \bibinfo {author} {\bibfnamefont {R.}~\bibnamefont {Oszwa{\l}dowski}},
  \bibinfo {author} {\bibfnamefont {T.}~\bibnamefont {Sato}}, \bibinfo {author}
  {\bibfnamefont {F.}~\bibnamefont {Matsukura}}, \bibinfo {author}
  {\bibfnamefont {T.}~\bibnamefont {Dietl}}, \bibinfo {author} {\bibfnamefont
  {H.}~\bibnamefont {Ohno}}, \ and\ \bibinfo {author} {\bibfnamefont
  {T.}~\bibnamefont {Takahashi}},\ }\bibfield  {title} {\enquote {\bibinfo
  {title} {{Fermi level position, Coulomb gap and Dresselhaus splitting in
  (Ga,Mn)As}},}\ }\href {\doibase 10.1038/srep27266} {\bibfield  {journal}
  {\bibinfo  {journal} {Sci. Rep.}\ }\textbf {\bibinfo {volume} {6}},\ \bibinfo
  {pages} {27266} (\bibinfo {year} {2016})},\ \Eprint
  {http://arxiv.org/abs/1606.02047} {arXiv:1606.02047} \BibitemShut {NoStop}%
\bibitem [{\citenamefont {Kumigashira}\ \emph {et~al.}(1998)\citenamefont
  {Kumigashira}, \citenamefont {Kim}, \citenamefont {Ito}, \citenamefont
  {Ashihara}, \citenamefont {Takahashi}, \citenamefont {Suzuki}, \citenamefont
  {Nishimura}, \citenamefont {Sakai}, \citenamefont {Kaneta},\ and\
  \citenamefont {Harima}}]{Kumigashira1998}%
  \BibitemOpen
  \bibfield  {author} {\bibinfo {author} {\bibfnamefont {H}~\bibnamefont
  {Kumigashira}}, \bibinfo {author} {\bibfnamefont {Hyeong-Do}\ \bibnamefont
  {Kim}}, \bibinfo {author} {\bibfnamefont {T}~\bibnamefont {Ito}}, \bibinfo
  {author} {\bibfnamefont {A}~\bibnamefont {Ashihara}}, \bibinfo {author}
  {\bibfnamefont {T}~\bibnamefont {Takahashi}}, \bibinfo {author}
  {\bibfnamefont {T}~\bibnamefont {Suzuki}}, \bibinfo {author} {\bibfnamefont
  {M}~\bibnamefont {Nishimura}}, \bibinfo {author} {\bibfnamefont
  {O}~\bibnamefont {Sakai}}, \bibinfo {author} {\bibfnamefont {Y}~\bibnamefont
  {Kaneta}}, \ and\ \bibinfo {author} {\bibfnamefont {H}~\bibnamefont
  {Harima}},\ }\bibfield  {title} {\enquote {\bibinfo {title} {{High-resolution
  angle-resolved photoemission study of LaSb}},}\ }\href {\doibase
  10.1103/PhysRevB.58.7675} {\bibfield  {journal} {\bibinfo  {journal} {Phys.
  Rev. B}\ }\textbf {\bibinfo {volume} {58}},\ \bibinfo {pages} {7675--7680}
  (\bibinfo {year} {1998})}\BibitemShut {NoStop}%
\bibitem [{\citenamefont {Kanski}\ \emph {et~al.}(2017)\citenamefont {Kanski},
  \citenamefont {Ilver}, \citenamefont {Karlsson}, \citenamefont {Ulfat},
  \citenamefont {Leandersson}, \citenamefont {Sadowski},\ and\ \citenamefont
  {{Di Marco}}}]{Kanski2017}%
  \BibitemOpen
  \bibfield  {author} {\bibinfo {author} {\bibfnamefont {J}~\bibnamefont
  {Kanski}}, \bibinfo {author} {\bibfnamefont {L}~\bibnamefont {Ilver}},
  \bibinfo {author} {\bibfnamefont {K}~\bibnamefont {Karlsson}}, \bibinfo
  {author} {\bibfnamefont {I}~\bibnamefont {Ulfat}}, \bibinfo {author}
  {\bibfnamefont {M}~\bibnamefont {Leandersson}}, \bibinfo {author}
  {\bibfnamefont {J}~\bibnamefont {Sadowski}}, \ and\ \bibinfo {author}
  {\bibfnamefont {I}~\bibnamefont {{Di Marco}}},\ }\bibfield  {title} {\enquote
  {\bibinfo {title} {{Electronic structure of (Ga,Mn)As revisited}},}\ }\href
  {\doibase 10.1088/1367-2630/aa5a42} {\bibfield  {journal} {\bibinfo
  {journal} {New J. Phys.}\ }\textbf {\bibinfo {volume} {19}},\ \bibinfo
  {pages} {023006} (\bibinfo {year} {2017})}\BibitemShut {NoStop}%
\bibitem [{\citenamefont {Bannow}\ \emph {et~al.}(2016)\citenamefont {Bannow},
  \citenamefont {Rubel}, \citenamefont {Badescu}, \citenamefont {Rosenow},
  \citenamefont {Hader}, \citenamefont {Moloney}, \citenamefont {Tonner},\ and\
  \citenamefont {Koch}}]{Bannow2016}%
  \BibitemOpen
  \bibfield  {author} {\bibinfo {author} {\bibfnamefont {Lars~C.}\ \bibnamefont
  {Bannow}}, \bibinfo {author} {\bibfnamefont {Oleg}\ \bibnamefont {Rubel}},
  \bibinfo {author} {\bibfnamefont {Stefan~C.}\ \bibnamefont {Badescu}},
  \bibinfo {author} {\bibfnamefont {Phil}\ \bibnamefont {Rosenow}}, \bibinfo
  {author} {\bibfnamefont {J{\"{o}}rg}\ \bibnamefont {Hader}}, \bibinfo
  {author} {\bibfnamefont {Jerome~V.}\ \bibnamefont {Moloney}}, \bibinfo
  {author} {\bibfnamefont {Ralf}\ \bibnamefont {Tonner}}, \ and\ \bibinfo
  {author} {\bibfnamefont {Stephan~W.}\ \bibnamefont {Koch}},\ }\bibfield
  {title} {\enquote {\bibinfo {title} {{Configuration dependence of band-gap
  narrowing and localization in dilute GaAs1-xBix alloys}},}\ }\href {\doibase
  10.1103/PhysRevB.93.205202} {\bibfield  {journal} {\bibinfo  {journal} {Phys.
  Rev. B}\ }\textbf {\bibinfo {volume} {93}},\ \bibinfo {pages} {1--10}
  (\bibinfo {year} {2016})}\BibitemShut {NoStop}%
\bibitem [{\citenamefont {Schmidt}\ and\ \citenamefont
  {Bechstedt}(1996)}]{Schmidt1996}%
  \BibitemOpen
  \bibfield  {author} {\bibinfo {author} {\bibfnamefont {W.~G.}\ \bibnamefont
  {Schmidt}}\ and\ \bibinfo {author} {\bibfnamefont {F.}~\bibnamefont
  {Bechstedt}},\ }\bibfield  {title} {\enquote {\bibinfo {title} {{Geometry and
  electronic structure of GaAs(001)(2x4) reconstructions.}}}\ }\href
  {http://www.ncbi.nlm.nih.gov/pubmed/9985804} {\bibfield  {journal} {\bibinfo
  {journal} {Phys. Rev. B}\ }\textbf {\bibinfo {volume} {54}},\ \bibinfo
  {pages} {16742--16748} (\bibinfo {year} {1996})}\BibitemShut {NoStop}%
\bibitem [{\citenamefont {Hogan}\ \emph {et~al.}(2003)\citenamefont {Hogan},
  \citenamefont {Paget}, \citenamefont {Garreau}, \citenamefont {Sauvage},
  \citenamefont {Onida}, \citenamefont {Reining}, \citenamefont {Chiaradia},\
  and\ \citenamefont {Corradini}}]{Hogan2003b}%
  \BibitemOpen
  \bibfield  {author} {\bibinfo {author} {\bibfnamefont {C.}~\bibnamefont
  {Hogan}}, \bibinfo {author} {\bibfnamefont {D.}~\bibnamefont {Paget}},
  \bibinfo {author} {\bibfnamefont {Y.}~\bibnamefont {Garreau}}, \bibinfo
  {author} {\bibfnamefont {M.}~\bibnamefont {Sauvage}}, \bibinfo {author}
  {\bibfnamefont {G.}~\bibnamefont {Onida}}, \bibinfo {author} {\bibfnamefont
  {L.}~\bibnamefont {Reining}}, \bibinfo {author} {\bibfnamefont
  {P.}~\bibnamefont {Chiaradia}}, \ and\ \bibinfo {author} {\bibfnamefont
  {V.}~\bibnamefont {Corradini}},\ }\bibfield  {title} {\enquote {\bibinfo
  {title} {{Early stages of cesium adsorption on the As-rich c(2x8)
  reconstruction of GaAs(001): Adsorption sites and Cs-induced chemical
  bonds}},}\ }\href {\doibase 10.1103/PhysRevB.68.205313} {\bibfield  {journal}
  {\bibinfo  {journal} {Phys. Rev. B}\ }\textbf {\bibinfo {volume} {68}},\
  \bibinfo {pages} {205313} (\bibinfo {year} {2003})}\BibitemShut {NoStop}%
\end{thebibliography}%

%{[}34{]} Kr¨omker B. et al. Development of a momentum microscope for
%time resolved band structure imaging. \emph{Review of Scientific
%Instruments} 79, 053702 (2008)

%\newpage
%\section{Supplementary information}

% %%%%%%%%%%%%%%%%%%%%%%%%%%%
% \begin{figure}
% \includegraphics[width=3.5in,height=6in]{Images/ARPES-L07-decapped.png}
% \caption{
% $k$-PEEM data of decapped 0.7\% Bi (right) in comparison with decapped GaAs (left) measured with a laboratory Helium lamp at $h\nu$ = 21.2~eV. (a) and (b) show respective $in situ$ LEED images revealing a 1x3 reconstruction of GaAsBi with respect to the GaAs(001) $2\times 4$ symmetry. 
% $k$-PEEM intensities along the directions [100], [0-10], [110] and [-110].
%  }
% \label{fig:kPEEM-decap-GaAsBi}
% \end{figure}
% %%%%%%%%%%%%%%%%%%%%%%%%%%%%%

%\includegraphics[width=5.53797in,height=3.62397in]{media/image9.emf}

%

% %%%%%%%%%%%%%%%%%%%%%%%%%%
% \begin{figure}
% \includegraphics[width=3.5in,height=2in]{Images/Fermi-level-after-Fe-deposition.png}
% \caption{Calibration of the GaAs(001) $2\times 4$ valence band top. (a) and (b) show full UPS scans in the binding energy range [-2.5, +18.5] eV and
% [-2.5, +5.3] eV, respectively, measured at a photon energy 21.2 eV
% in the initial state and after Fe deposition (black and red lines). In
% (c) an ARPES image of GaAs [100] cut is shown for direct comparison
% in the binding energy range.}
% \label{fig:Fermi-level-NanoESCA}
% \end{figure}
% %%%%%%%%%%%%%%%%%%%%%%%%%%%%%

%
%%%%%%%%%%%%%%%%%%%%%%%%%%
\begin{figure}
\includegraphics[width=3.2in,height=3.5in]{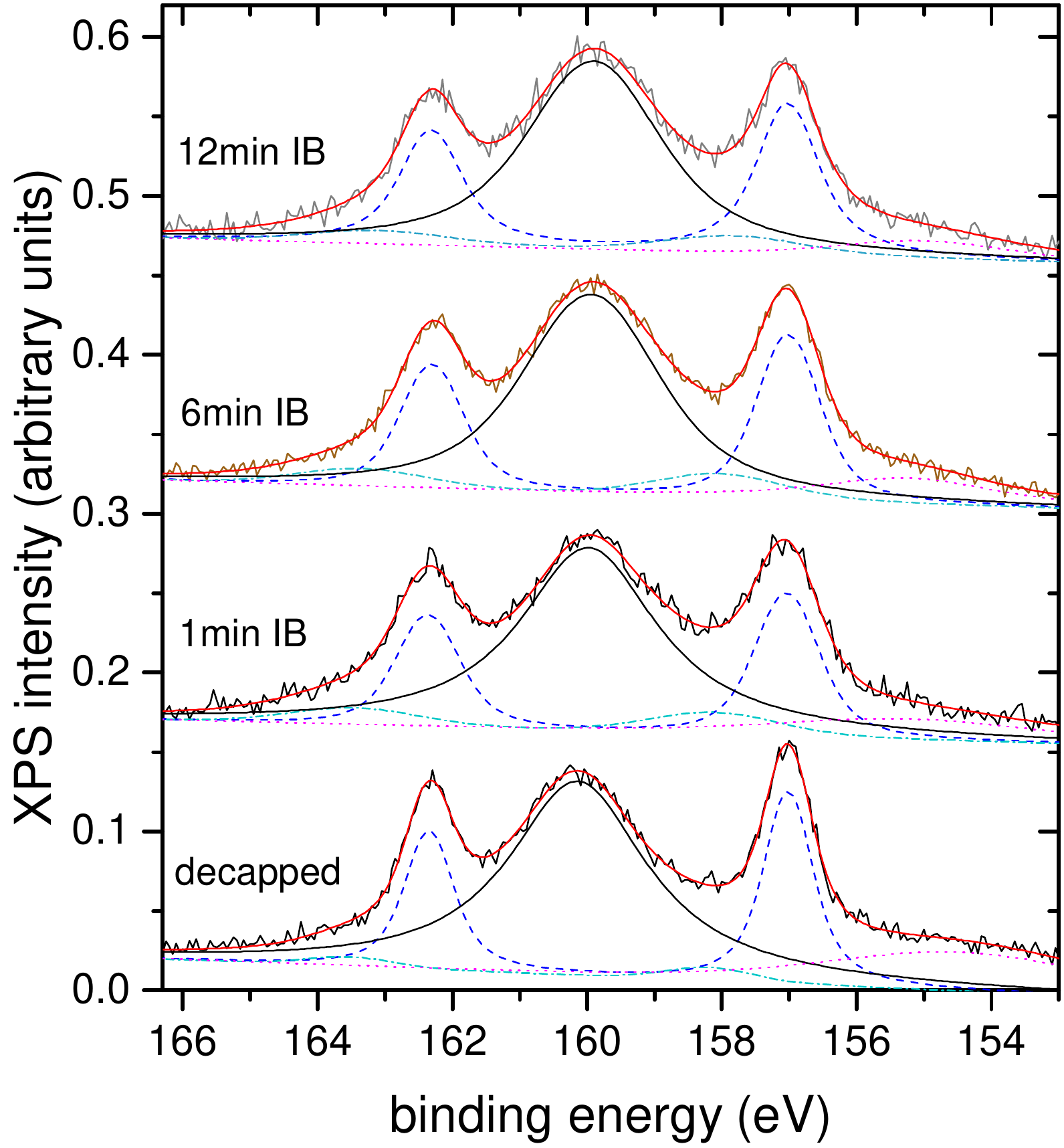}
\caption{Supplementary material: Effect of sputtering on the Bi 4f XPS data. Bottom to top show the sample in different stages starting from the initial decapped state and after increasing sputtering times.}
\label{fig:sputtering-effect-SI}
\end{figure}
%%%%%%%%%%%%%%%%%%%%%%%%%%%%%

\end{document}